\PassOptionsToPackage{table}{xcolor}
\documentclass[10pt]{article}
\usepackage[total={6in,9.5in}]{geometry}
\usepackage{amsmath,amssymb,amsthm,booktabs}
\usepackage{siunitx}
\usepackage{optidef}
\usepackage{palatino}
\usepackage{graphicx,epstopdf,epsfig}
\usepackage[dvipsnames]{xcolor}
\usepackage{float}
\usepackage[caption=false,font=footnotesize]{subfig}
\usepackage{url}
\usepackage{comment}
\usepackage{microtype}
\usepackage[normalem]{ulem}
\usepackage[section]{placeins} 
\usepackage{centernot}
\usepackage{enumitem}
\usepackage[noadjust]{cite}
\usepackage{hyperref}
\usepackage{cleveref}
\usepackage{xcolor}  
\usepackage[table]{xcolor}

\usepackage{colortbl}
\usepackage{caption}
\usepackage{subfig}
\usepackage{algorithmic}
\usepackage{algorithm}
\newtheorem{thm}{Theorem}
\newtheorem{prop}{Proposition}

\newtheorem{lemma}{Lemma}

\def\Re{\mathbb{R}}
\def\R{\mathbf{R}}
\def\K{\mathbf{K}}
\def\X{\mathbf{X}}
\def\A{\mathbf{A}}
\def\B{\mathbf{B}}
\def\S{\mathbf{S}}
\def\Z{\mathbf{Z}}

\def\M{\mathbf{M}}

\def\Y{\mathbf{Y}}
\def\E{\mathbf{E}}
\def\H{\matcal{H}}
\def\H{\mathbf{H}}

\def\bbH{\mathbb{H}}
\def\W{\mathbf{W}}

\def\cG{\mathcal{G}}
\def\I{\mathbf{I}}
\def\D{\mathbf{D}}
\def\F{\mathbf{F}}
\def\cL{\mathcal{L}}
\def\cK{\mathcal{K}}
\def\Q{\mathbf{Q}}
\def\cD{\mathcal{D}}
\def\cP{\mathcal{P}}
\def\cT{\mathcal{T}}
\def\cW{\mathcal{W}}
\def\cV{\mathcal{V}}

\newcommand{\tr}{\mathrm{tr}}
\newcommand{\x}{\boldsymbol{x}}
\newcommand{\y}{\boldsymbol{y}}
\newcommand{\e}{\boldsymbol{e}}
\newcommand{\p}{\boldsymbol{p}}

\newcommand{\0}{\mathbf{0}}
\newcommand{\fix}{\mathrm{fix}\,}

\DeclareMathOperator*{\diag}{diag}

\begin{document}

\title{HyDeFuse: Provably Convergent Denoiser-Driven Hyperspectral Fusion}

\date{}

\author{%
Sagar Kumar$^{1}$, 
Unni V. S.$^{2}$, 
Kunal N. Chaudhury$^{3,*}$\\[6pt]
\begin{minipage}[t]{0.9\textwidth}
\centering
\small
$^{1}$Department of Electrical Engineering, Indian Institute of Science, Bengaluru, India\\
$^{2}$Staff Data Scientist, Walmart Global Tech, India\\
$^{3}$Department of Electrical Engineering, Indian Institute of Science, Bengaluru, India\\[4pt]
Email: \href{mailto:sagarkumar@iisc.ac.in}{sagarkumar1@iisc.ac.in}, 
\href{mailto:unni@iisc.ac.in}{unniv@alum.iisc.ac.in}, 
\href{mailto:kunal@iisc.ac.in}{kunal@iisc.ac.in}\\[4pt]
$^{*}$\textit{Corresponding author:} \href{mailto:kunal@iisc.ac.in}{kunal@iisc.ac.in}
\end{minipage}%
}

\footnotetext[3]{K.~N.~Chaudhury was supported by the ANRF, Government of India, under grant STR/2021/000011}


\maketitle

\begin{abstract}
Hyperspectral (HS) images provide fine spectral resolution but have limited spatial resolution, whereas multispectral (MS) images capture finer spatial details but have fewer bands. HS-MS fusion aims to integrate HS and MS images to generate a single image with improved spatial and spectral resolution. This is commonly formulated as an inverse problem with a linear forward model. However, reconstructing high-quality images using the forward model alone is challenging, necessitating the use of regularization techniques. In this work, we investigate the paradigm of denoiser-driven regularization, where a powerful off-the-shelf denoiser is used for implicit regularization within an
iterative algorithm. This has shown much promise but remains relatively underexplored in hyperspectral imaging. The technical challenge lies in designing hyperspectral denoisers that can guarantee convergence—while strong denoisers can produce high-quality reconstructions, they may also cause instability or divergence. Specifically, we consider a denoiser-driven fusion algorithm, HyDeFuse, which leverages a class of pseudo-linear denoisers for implicit regularization. We demonstrate how the contraction mapping theorem can be applied to establish global linear convergence of HyDeFUse. Finally, we validate our theoretical findings and present fusion results on publicly available datasets to demonstrate the performance of HyDeFuse.
\end{abstract}
\vspace{1em}


\section{Introduction}

Hyperspectral cameras capture images of a scene over a continuous range of spectral bands. Because the incoming light is divided into many narrow bands, the energy in each band decreases, resulting in an inherent trade-off between spatial and spectral resolution. Hyperspectral (HS) images typically offer high spectral resolution (numerous bands) but suffer from low spatial resolution. In contrast, multispectral (MS) images capture high spatial resolution but contain limited spectral information. HS-MS fusion techniques combine HS and MS images of the same scene to generate an image with both high spatial and spectral resolution~\cite{dian2021recent,vivone2023multispectral}. Such fused images are useful 
in applications such as navigation, surveillance, object identification, medical diagnosis, and classification~\cite{prasad2020hyperspectral,huang2024spatial,li2024feedback}.

HS-MS fusion faces several practical challenges, such as misregistrations, atmospheric changes, and inconsistent illumination~\cite{zhou2019integrated}. Most studies simplify the problem by assuming perfect registration and stable conditions. However, even under these ideal settings, producing high-quality fused images requires advanced reconstruction algorithms.

\subsection{Literature review}

HS-MS fusion techniques can be categorized into five groups: pansharpening~\cite{vivone2023multispectral,aiazzi2007improving}, Bayesian~\cite{simoes2014convex,eismann2004resolution,dong2022spatial,liu2022patch,bungert2018blind}, multiresolution analysis~\cite{aiazzi2006mtf}, tensor factorization~\cite{dian2017hyperspectral,dian2019nonlocal,fang2024cs2dips,yokoya2011coupled,xu2024coupled,zhang2024hyperspectral}, and deep learning~\cite{ciotola2022pansharpening,xie2019multispectral,wang2021enhanced,dian2021recent,gao2022hyperspectral}. Of these, Bayesian, tensor factorization, and deep learning have garnered attention due to their state-of-the-art performance. In Bayesian methods, the fusion problem is formulated as a maximum a posteriori (MAP) optimization, incorporating appropriate priors on the ground truth. Commonly used priors include Gaussian priors~\cite{wei2015bayesian}, Markov Random Fields~\cite{joshi2009map}, transform-domain sparsity priors~\cite{bungert2018blind,zhang2024hyperspectral,simoes2014convex}, and low-rank and nonnegativity priors~\cite{lin2022hyperspectral}. On the other hand, deep learning methods attempt to learn the functional relationship between the ground truth and the observed HS-MS images using end-to-end deep networks. They can achieve remarkable results, leveraging architectures such as transformers~\cite{jia2023multiscale, hu2022fusformer, ma2024reciprocal, cao2024unsupervised, fang2024mimo}, CNNs~\cite{ge2020hyperspectral, ran2023guidednet, cai2024atsfcnn}, and diffusion models~\cite{li2023hyperspectral, yu2024unmixdiff, wu2023hsr, li2025kandiff}. Unfortunately, deep learning methods require large training datasets, which are often hard to obtain when high-resolution ground truth images are unavailable~\cite{dian2021recent}.
Moreover, training deep models is computationally demanding, and models trained for specific forward models often lack adaptability~\cite{dian2021recent}.

\begin{figure}[t!]
\centering
  \begin{minipage}{0.5\linewidth}
    \centering
    \includegraphics[width=\linewidth]{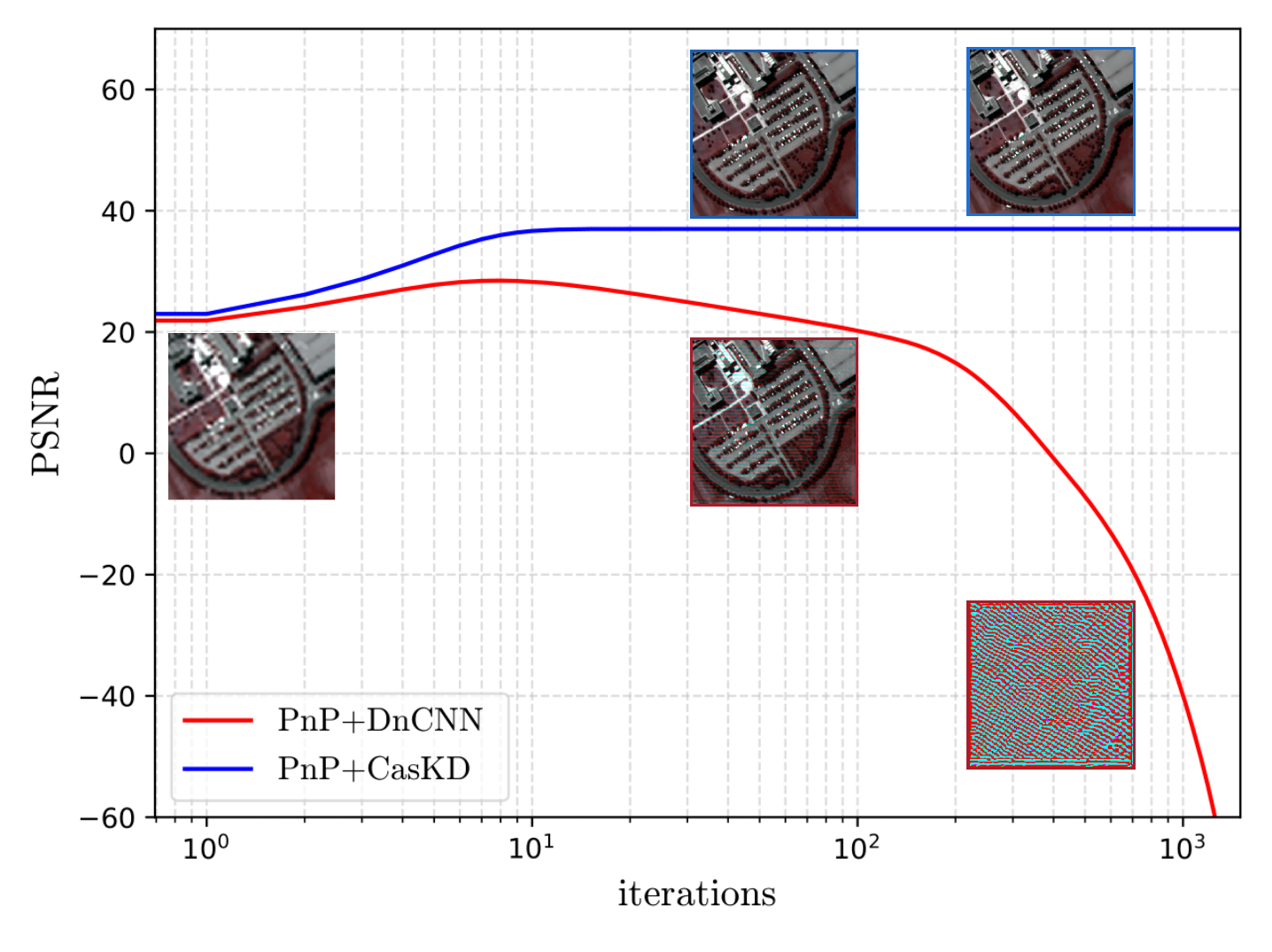}
  \end{minipage}%
  \begin{minipage}{0.5\linewidth}
    \centering
    \includegraphics[width=\linewidth]{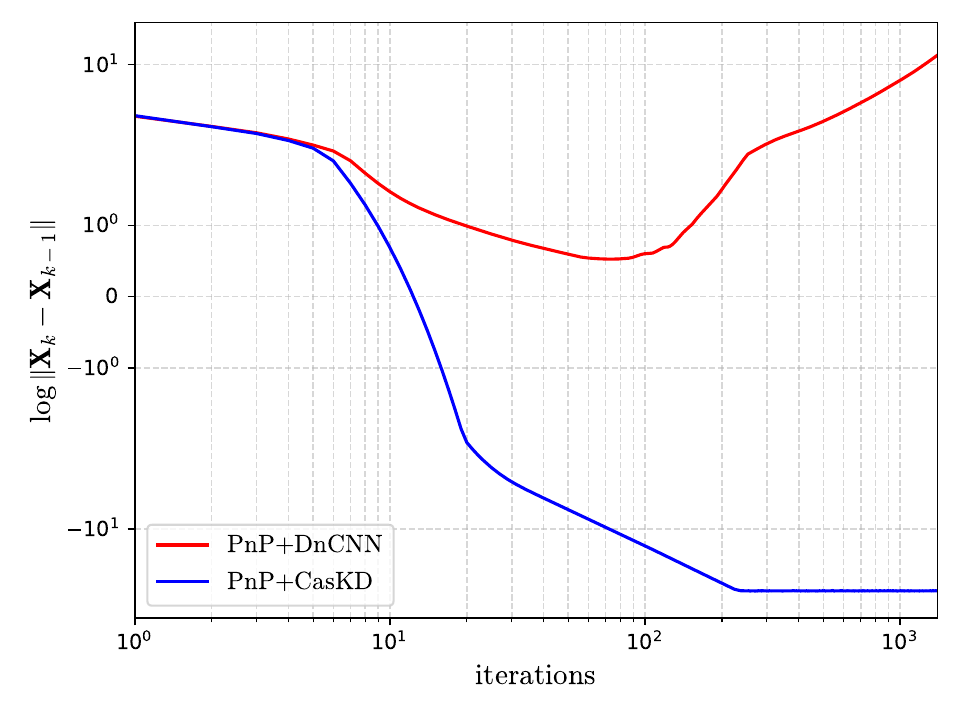}
  \end{minipage}
  \caption{An illustration of instability in denoiser-driven fusion on the Pavia dataset. The inset images show the fused results at iterations 1, 100, and 1500. Specifically, we compare the performance of Proximal Gradient Descent (PnP-PGD) using DnCNN~\cite{zhang2017beyond} and the proposed CasKD denoiser. The plots show the evolution of PSNR and the distance between successive iterates  $\{\X_k\}$. The PSNR evolution serves as an indicator of stability: ideally, it should improve gradually rather than rise initially and then degrade after many iterations. Notably, with DnCNN, the PSNR initially improves but then starts dropping after a certain point, ultimately leading to a poor reconstruction. The instability is also highlighted in the second plot, showing the divergence of the iterates. This phenomenon is common to deep denoisers and has been previously observed for other inverse problems~\cite{terris2024equivariant}. In contrast, the PSNR for our denoiser increases steadily. Moreover, the successive differences decay to zero, which aligns with the theoretical guarantees established in this work.}
  \label{fig:div}
\end{figure}

\subsection{Motivation}

HS–MS fusion is closely related to the classical problem of image superresolution in computational imaging. Recently, denoisers have been successfully incorporated into such imaging tasks, yielding state-of-the-art results~\cite{romano2017little,sreehari2016plug,dian2020regularizing,wang2023deep}. Our work focuses on the Plug-and-Play (PnP) method, where an off-the-shelf denoiser is used as a implicit regularizer inside a classical reconstruction algorithm. The denoiser acts as an prior, guiding the iterative process towards a better reconstruction. In end-to-end approaches, the forward model is incorporated during training, which means that any change to the forward model necessitates retraining.  In contrast, PnP allows the decoupling of the forward model and the regularizer,  so that the same pretrained denoiser can be used across different models. The choice of the denoiser in PnP can significantly impact the reconstruction quality. Classical denoisers~\cite{LLRT,kbr,nair2019hyperspectral,NGMeet,zhuang2021fasthymix,he2020non,tu2023new} do not require training and offer better interpretability and generalizability. Deep denoisers have recently gained popularity due to their strong denoising capability~\cite{zhang2017beyond,lin2019hyperspectral,cao2021deep,NGMeet}. In the PnP framework, the denoiser functions as an implicit regularizer or smoothing operator rather than serving solely as a noise-removal tool~\cite{nair2021fixed,gavaskar2021plug}.

The flip side is that PnP algorithms are not strictly derived from an optimization formulation, and hence the convergence of their iterates cannot be guaranteed automatically. It is thus not surprising that the convergence of PnP methods has been an active area of research~\cite{hertrich2021convolutional,sherry2024designing,bredies2024learning,sreehari2016plug, teodoro2018convergent,gavaskar2021plug, cohen2021regularization,ryu2019plug, athalye2023contractivity}. Unlike traditional denoising, where the denoiser is applied once to a noisy image, PnP algorithms employ the denoiser iteratively. Consequently, even a relatively weak denoiser can be an effective regularizer, producing a high-quality reconstruction. Instead, the challenge lies in ensuring that the PnP iterations remain stable and do not diverge from their intended path. By “stable,” we refer to the consistency of PSNR, which should improve steadily rather than rise initially and then degrade after many iterations. As shown in Figure~\ref{fig:div}, powerful deep denoisers such as DnCNN~\cite{zhang2017beyond} can exhibit this type of instability, with the iterates diverging after a certain point. We have also observed similar behavior in hyperspectral imaging, highlighting the need for denoisers that ensure convergence or, at the very least, maintain PSNR stability.  In this context, it is worth noting that existing convergence guarantees for deep denoisers typically rely on strong technical assumptions, which are often impractical or difficult to verify in real-world settings.

\subsection{Contribution}

Recent works have demonstrated that a class of pseudo-linear denoisers, known as kernel denoisers, can be used to develop convergent PnP algorithms~\cite{nair2019hyperspectral,athalye2023contractivity,sreehari2016plug,gavaskar2020plug}. The convergence guarantee stems from the unique mathematical properties of kernel denoisers, which are challenging to establish for (black-box) deep denoisers. While kernel denoisers are not as powerful as DnCNN \cite{zhang2017beyond} and DRUNET \cite{devalla2018drunet}, they have been shown to yield good reconstructions for inverse problems such as deblurring and superresolution~\cite{zhang2021plug,nair2019hyperspectral,sreehari2016plug,ryu2019plug}. Kernel denoisers have also been used for convergent hyperspectral fusion~\cite{nair2019hyperspectral}. However, as explained in Section~\ref{sec:analysis}, the analysis in these works does not apply to our PnP algorithm. This is because we utilize an enhanced denoiser that lacks specific properties (such as symmetry) required in previous analyses. Our contributions are as follows:

\begin{enumerate}

\item \textbf{Cascaded denoiser}. We introduce a hyperspectral denoiser, CasKD, by cascading two existing kernel denoisers~\cite{nair2019hyperspectral,sreehari2016plug}. The first denoiser is designed to capture inter-band correlations, leveraging spectral dependencies in hyperspectral images, while the second denoiser exploits the in-band correlations. We demonstrate that CasKD provides superior denoising and fusion capabilities compared to either denoiser alone, although at an increased computational cost. This improvement is significant, as simply cascading two denoisers does not always ensure better performance. We also analyze the mathematical properties of CasKD that are necessary for the convergence analysis.
    
\item \textbf{Convergence analysis}. We integrate the proposed CasKD denoiser with the proximal gradient descent algorithm~\cite{beck2017first} to develop a fusion algorithm called HyDeFuse. We prove that HyDeFuse is guaranteed to converge linearly to a unique reconstruction for any arbitrary initialization. To the best of our knowledge, there are not many (iterative) fusion algorithms where the iterates are guaranteed to converge at a geometric rate. Importantly, the analysis enables us to determine the optimal step size required for convergence.

\item  \textbf{Validation and comparison}. We validate our theoretical results and conduct extensive experiments on publicly available datasets to demonstrate that HyDeFuse is competitive with state-of-the-art fusion methods.

\end{enumerate}

\subsection{Organization}

We review background materials in Section~\ref{sec:back}, including the forward model and the loss function for HS-MS fusion. The proposed denoiser and the fusion algorithm are presented in Section~\ref{sec:algo}, where we also discuss their mathematical properties. The core part is Section~\ref{sec:analysis}, where we analyze the convergence of HyDeFuse. Finally, in Section~\ref{sec:results}, we validate our theoretical findings and compare the performance of HyDeFuse with existing fusion algorithms. Proofs of technical results are provided in \Cref{appendix}.

\section{Background}
\label{sec:back}

\subsection{Hyperspectral Imaging}

The problem of hyperspectral–multispectral (HS–MS) fusion can be posed as an inverse problem governed by a linear forward model. This formulation arises from simplified assumptions about the imaging optics, which in reality involve nonlinear distortions, sensor noise, and optical aberrations~\cite{thomas2008synthesis,aly2014regularized}. In hyperspectral imaging, the sensor plane comprises an array of detectors that record the intensity of incoming light at various wavelengths. It is often modeled as a two-dimensional lattice, where sensors are positioned on a uniform sampling grid~\cite{aly2014regularized}.

In MS imaging, each sensor is responsive to a specific region of the electromagnetic spectrum. For a given ground-truth resolution, MS sensors are more densely distributed, resulting in a higher spatial sampling rate than in HS imaging. Consequently, MS images typically offer superior spatial resolution. However, since each MS pixel integrates information from multiple spectral bands, this comes at the cost of spectral degradation.

In contrast, HS sensors capture light over several narrow and contiguous spectral bands. Although this enables fine spectral discrimination across a wide range, the lower sensor density and narrower bands lead to spatial averaging at the detector plane, making HS images appear as blurred representations of the high-resolution scene. Additionally, the reduced sampling rate and limited energy per band result in spatial downsampling at the sensor level.

\subsection{Forward Model}

The above imaging process can be modeled as a combination of low-pass filtering and downsampling applied to an unknown  high-resolution image with fine spatial and spectral details~\cite{aly2014regularized,simoes2014convex}. The corresponding inverse problem aims to recover this high-resolution image from the observed HS and MS measurements. 

The unknown image is naturally represented as a 3-D tensor, whose dimensions correspond to spatial rows, spatial columns, and spectral bands. For computational convenience, this 3-D tensor is often reshaped into a matrix, where each column corresponds to the vectorized image of a specific spectral band. Under this convention, the standard forward model can be expressed as
\begin{eqnarray}
    \label{eq:fullmodel}
  \Y_{h} = \A\Z + \mathbf{\Theta}_h, \qquad  \Y_{m} = \Z\R  + \mathbf{\Theta}_m,
\end{eqnarray} 
where

\begin{itemize}

\item[(i)] \( \Z \in \Re^{N_m \times L_h} \) represents the ground truth image with high spectral and spatial resolution, where \( N_m \) is the number of pixels and \( L_h \) is the number of spectral bands.
        
\item[(ii)] \( \A = \S \B \) represents spatial degradation, where \( \S \in \Re^{N_h \times N_m} \) represents spatial subsampling and \( \B \in \Re^{N_{m} \times N_{m}} \) represents spatial blurring,
    \item[(iii)] \( \R \in \Re^{L_h \times L_m} \) represents spectral degradation, with each column representing the spectral response of a specific band,
      
\item[(iv)] \( \Y_h \in \Re^{N_h \times L_h} \) and \( \Y_m \in \Re^{N_m \times L_m} \) are the observed HS and MS images, and 

    \item[(v)] \( \mathbf{\Theta}_{h} \in \Re^{N_h \times L_h} \) and \(\mathbf{\Theta}_{m} \in \Re^{N_m \times L_m} \) represent  white Gaussian noise.
\end{itemize}

In the above representation, each column of \( \Z \) corresponds to the vectorized form of a spectral band, and the matrices $\B,\S$ (resp. $\R$) act on the columns (resp. rows) of $\Z$. In practice, the matrices $\B$, $\S$, and $\R$ are not stored explicitly but are implemented as linear operators.

\subsection{Variable Reduction}

The simplified forward model in~\eqref{eq:fullmodel} reduces the fusion task to solving a system of linear equations. The total number of equations and unknowns in~\eqref{eq:fullmodel} are $N_h L_h + N_m L_m$ and $N_m L_h$, respectively. Typically, $N_h \ll N_m$ and $L_m \ll L_h$, implying that the number of unknowns exceeds the number of measurements. As a result, estimating $\Z$ becomes a highly ill-posed problem~\cite{nair2019hyperspectral}, thereby necessitating regularization.

To further reduce the number of variables, we follow prior work~\cite{simoes2014convex,wei2015fast} and employ dimensionality reduction. The key assumption is that the spectral bands in $\Z$ are highly correlated, and hence $\Z$ lies in a low-dimensional subspace. Mathematically, this can be expressed as
\begin{equation}
\label{eq:decomp}
\Z = \X \E, \qquad \X \in \Re^{N_m \times L_s}, \ \E \in \Re^{L_s \times L_h},
\end{equation}
where the rows of $\E$ form a basis for the low-dimensional subspace. Substituting~\eqref{eq:decomp} into~\eqref{eq:fullmodel} yields
\begin{equation}
\label{eq:model}
\Y_h = \A \X \E + \mathbf{\Theta}_h, \qquad
\Y_m = \X \E \R + \mathbf{\Theta}_m,
\end{equation}
where $\X$ is the variable to be estimated. The final reconstruction is then given by $\widehat{\Z} = \widehat{\X} \E$, where $\widehat{\X}$ denotes the solution of~\eqref{eq:model}.

A standard approach for estimating $\E$ from the observed hyperspectral image $\Y_h$ is to use singular value decomposition~(SVD)~\cite{dian2020regularizing,simoes2014convex}. Specifically, $\Y_h$ is first upsampled to match the spatial resolution of $\Z$, and the resulting image, denoted by $\Y$, serves as a surrogate for $\Z$. We then compute its SVD as
\begin{equation*}
\Y = \sum_{j=1}^{r} \sigma_j \boldsymbol{u}_j \boldsymbol{v}_j^\top,
\end{equation*}
where $r$ is the rank of $\Y$, $\sigma_1 \geqslant \cdots \geqslant \sigma_r$ are the singular values, and $\boldsymbol{u}_1, \ldots, \boldsymbol{u}_r \in \Re^{N_m}$ and $\boldsymbol{v}_1, \ldots, \boldsymbol{v}_r \in \Re^{L_h}$ are the left and right singular vectors, respectively. 

Following the approach in~\cite{simoes2014convex}, we select a target subspace dimension $L_s \leqslant r$ and define
\begin{equation*}
\E = \big[ \boldsymbol{v}_1 \ \cdots \ \boldsymbol{v}_{L_s} \big]^\top \in \Re^{L_s \times L_h},
\end{equation*}
which serves as the basis of the low-dimensional subspace in~\eqref{eq:model}. The dimension $L_s$ is chosen such that most of the signal energy is preserved after projection onto the subspace. If $L_s$ is too small, important spectral information may be lost. In practice, we found that the performance is not highly sensitive to the exact value of $L_s$. A smaller $L_s$ also reduces computational cost per iteration, since fewer spectral bands are processed. Hence, it is preferable to choose the smallest $L_s$ that still retains most of the signal energy. The specific choice of $L_s$ is discussed in~\Cref{subsec:para}.

\subsection{Definitions and Notations} 

We use bold lowercase letters ($\x$, $\e$, etc.) to denote vectors, bold uppercase letters (e.g., $\X$, $\Z$) to denote matrices, and calligraphic letters ($\cT$, $\cW$, etc.) to denote operators acting on matrices.
For any $ \x \in \Re^n $, $\diag(\x) \in \Re^{n \times n}$ denotes the diagonal matrix with $\x$ on its diagonal. The identity matrix is denoted by $\I$, with its dimension understood from the context.

We use $\bbH=\Re^{N_m \times L_s}$ to denote the space of the matrix-valued variable $\X$ in \eqref{eq:model}. The reconstruction algorithm operates in this vector space. To analyze convergence, we equip $\bbH$ with an inner product and its induced norm. Specifically, we use the inner product
\begin{equation}
\label{eq:innerprod}
\langle \X_1, \X_2 \rangle_{\bbH} = \tr(\X_1^\top \X_2), \qquad \X_1, \X_2 \in \bbH,
\end{equation}
where $\tr$ denotes the trace operator. The norm induced by this inner product is given by 
\begin{equation*}
\| \X \|_{\mathbb{H}} = \sqrt {\langle \X, \X \rangle}_{\bbH} \qquad (\X \in \bbH).
\end{equation*}

Since the columns of $\X \in \bbH$ represent spectral bands, the inner product~\eqref{eq:innerprod} amounts to computing the standard dot product between the corresponding bands of $\X_1$ and $\X_2$, followed by summation over all bands. In particular, 
\begin{equation*}
\| \X \|^2_{\mathbb{H}} = \sum_{i,j} |\X_{ij}|^2,
\end{equation*}
which is just the squared Frobenius norm of $\X$.

We use $\cL(\bbH)$ to denote the class of linear operators that map $\bbH$ into $\bbH$. We use $\sigma(\M)$ and $\sigma(\cT)$ for the spectrum (set of eigenvalues) of a matrix $\M$ or an operator $\cT$. We use $\sigma_{\max}(\M)$ for the largest singular value of $\M$. 

We say that $\X$ is a fixed point of an operator $\cT$ if $\cT(\X) = \X$. The set of all fixed points of $\cT$ is denoted by $\fix(\cT)$. If $\cT$ is linear, then $\fix(\cT)$ coincides with the eigenspace corresponding to the eigenvalue $1$.

We will need the concept of a self-adjoint operator, which generalizes the idea of a symmetric matrix~\cite{meyer2023matrix}. An operator $\cT$ on $(\bbH, \langle \cdot, \cdot \rangle_{\bbH})$ is said to be self-adjoint if
\begin{equation*}
\langle \cT(\X_1), \X_2 \rangle_{\bbH} = \langle \X_1, \cT(\X_2) \rangle_{\bbH}, \qquad \X_1, \X_2 \in \bbH.
\end{equation*}
A self-adjoint operator (or a symmetric matrix) is called positive semidefinite if all its eigenvalues are nonnegative.

 An operator $\cT : \bbH \to \bbH$ is said to be $L$-Lipschitz if there exists $L> 0$ such that
\[
\|\cT(\X_1) - \cT(\X_2)\|_{\bbH}
\leqslant L \|\X_1 - \X_2\|_{\bbH},
\qquad  (\X_1, \X_2 \in \bbH).
\]
In particular, $\cT$ is a contraction operator if $L < 1$, and a nonexpansive operator if $L = 1$. A matrix $\M$ is nonexpansive if and only if $\sigma_{\max}(\M) \leqslant 1$.

A function $\ell: \bbH \to \Re$ is said to be convex if $\ell(\theta \X_1+ (1-\theta) \X_2) \leqslant \theta \ell( \X_1)+ (1-\theta) \ell(\X_2)$ for all $\theta \in (0,1)$ and $\X_1,\X_2 \in \bbH$.  Moreover, $\ell$ is said to be $\beta$-smooth if it is differentiable and its gradient operator $\nabla \!\ell: \bbH \to \bbH$ is $\beta$-Lipschitz.

A matrix $\W$ is called stochastic if all its entries are nonnegative and each row sums to $1$. It is said to be irreducible if, for every pair of indices $i$ and $j$, there exists $k \geqslant 1$ such that $(\W^k)_{ij} > 0$. Equivalently, the directed graph associated with $\W$ is strongly connected~\cite{meyer2023matrix}.

\section{Fusion Algorithm}
\label{sec:algo}

\subsection{Plug-and-Play Framework}

\begin{figure}[t]
  \centering
  \begin{minipage}{1\linewidth}
    \centering
\includegraphics[width=\linewidth]{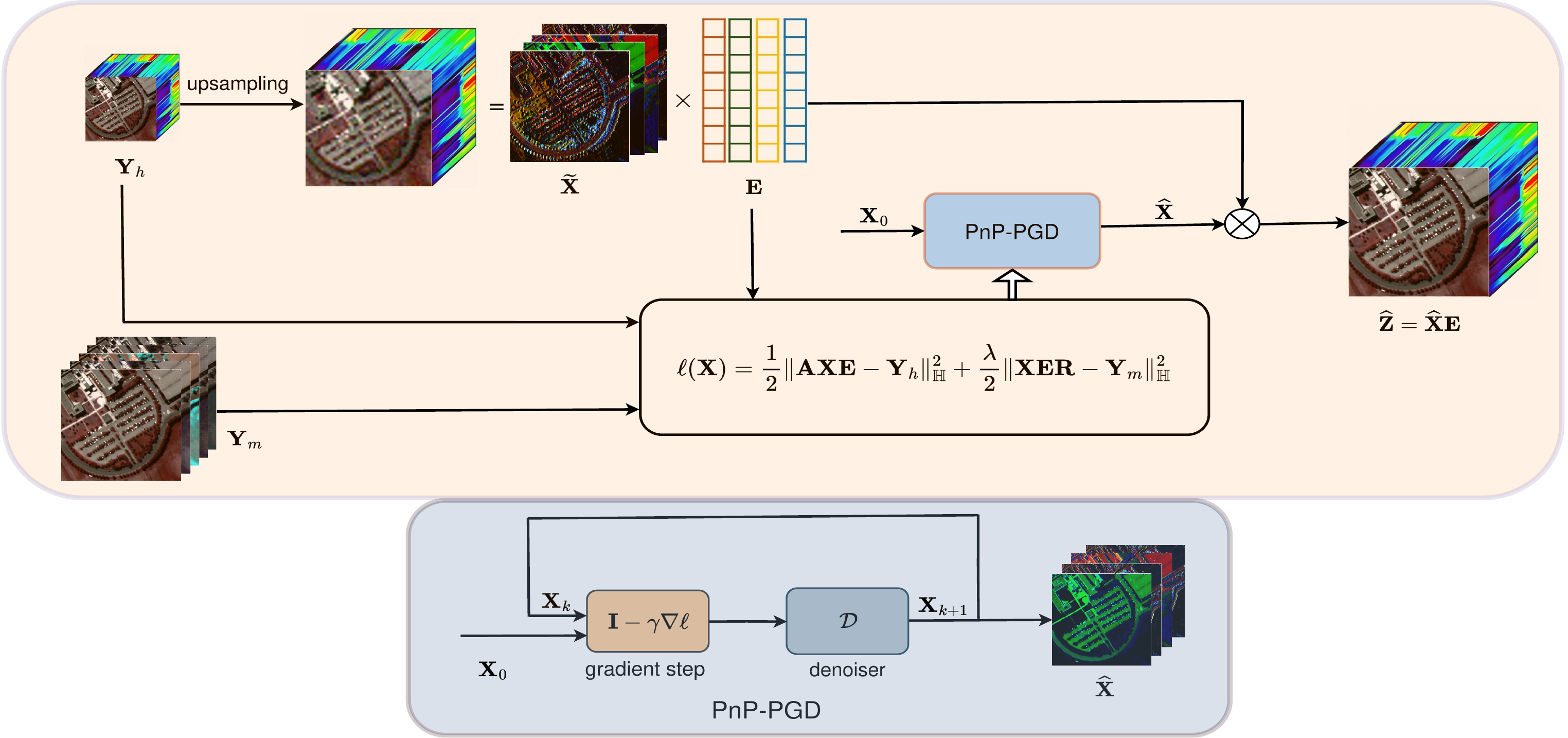}
  \end{minipage}
  \caption{The processing pipeline of HyDeFuse, the proposed fusion algorithm. The HS image $\Y_h$ is first upscaled (interpolated) to match the spatial resolution of the MS image, which is then used to estimate the spectral subspace $\E$. The loss function $\ell$ is formulated using $\E$, the observed HS and MS images ($\Y_h$ and $\Y_m$), and the forward operators $\A$ and $\R$. The reconstruction is performed using the iterative PnP-PGD algorithm, where each iteration alternates between a gradient step on $\ell$ and regularization via the proposed hyperspectral denoiser $\cD$. The output of PnP-PGD is $\widehat{\X}$ and $\widehat{\Z} = \widehat{\X} \E$ is the output of HyDeFuse.
  }
  \label{fig:Block_daigram}
\end{figure}

We begin by introducing the Plug-and-Play (PnP) framework, which forms the foundation of the proposed fusion algorithm. The pipeline of our method is illustrated in Figure \ref{fig:Block_daigram}. The design and properties of the kernel-based denoiser~$\cD$ are discussed in the following subsection.

The PnP framework~\cite{sreehari2016plug} originates from the classical optimization formulation for solving~\eqref{eq:model}:
\begin{equation}
\label{eq:fg}
\underset{\X \in \bbH}{\min} \ \ \ell(\X) + \varphi(\X), 
\end{equation}
where the model-based loss function $\ell : \bbH \to \Re$ is given by
\begin{equation}
\label{eq:loss}
\ell(\X) = \frac{1}{2} \| \A \X \E - \Y_h \|_{\bbH}^2 + \frac{\lambda}{2} \| \X \E \R - \Y_m \|_{\bbH}^2,
\end{equation}
 and $\varphi: \bbH \to \Re$ is a (convex) regularizer. Since $\varphi$ is typically non-smooth, proximal algorithms are used to perform the optimization~\cite{aly2014regularized,simoes2014convex,zhang2018pet,kazantsev2018joint}. The loss $\ell$ is convex, as it is the sum of convex quadratic functions. 
 
Since $\ell$ is differentiable, the proximal gradient descent (PGD) algorithm can be used to solve the optimization problem~\eqref{eq:fg}. For a given initialization \( \X_0 \in \bbH \), the PGD iterations are given by
 \begin{equation}
\label{eq:pgd}
\X_{k+1} = \mathrm{prox}_{\gamma \varphi} \big(\X_k - \gamma \nabla \! \ell(\X_k) \big) \qquad (k \geqslant 0),
\end{equation}
where $\gamma >0$ is the step size, $\nabla \! \ell$ is the gradient of $\ell$, and $\mathrm{prox}_{\gamma \varphi}: \bbH \to \bbH$ is the proximal operator of $\gamma \varphi$ given by
\begin{equation}
\label{eq:prox}
\mathrm{prox}_{\gamma \varphi} (\X) =  \underset{\H \in \bbH}{\mbox{argmin}} \ \frac{1}{2 } \|\H - \X\|_{\bbH}^2+ \gamma \varphi(\H).
\end{equation}

Instead of relying on a classical regularizer and its proximal operator, which typically acts as a Gaussian denoiser~\cite{sreehari2016plug}, the key innovation in the Plug-and-Play (PnP) framework is to replace the proximal operator in~\eqref{eq:pgd} with a generic denoising operator $\cD: \bbH \to \bbH$. Accordingly, the PnP update is given by
\begin{equation}
\label{eq:pnppgd}
\X_{k+1} = \cD \big(\X_k - \gamma \nabla \ell(\X_k)\big), \qquad k = 0, 1, \ldots
\end{equation}
The motivation behind this substitution is that denoisers such as DnCNN~\cite{zhang2017beyond,dian2020regularizing}, and DRUNet~\cite{devalla2018drunet} are often far more effective than the proximal operators in~\eqref{eq:prox} derived from classical regularizers such as total variation, wavelet sparsity, or low-rank models~\cite{simoes2014convex}. The expectation is that employing a stronger denoiser $\cD$ in~\eqref{eq:pnppgd} leads to higher-quality reconstructions.

The main technical challenge, however, is that the transition from~\eqref{eq:pgd} to~\eqref{eq:pnppgd} is not theoretically justified, with no convergence guarantee on the sequence $\{\X_k\}$ produced by~\eqref{eq:pnppgd}. Indeed, as illustrated in Figure~\ref{fig:div},  the iterative process can sometimes diverge. Hence, it is essential to design a denoiser $\cD$ that not only performs well empirically but also promotes convergence of the iterative scheme.

In the rest of the paper, we refer to~\eqref{eq:pnppgd} as PnP-PGD. This algorithm is not new and has been used in previous PnP studies~\cite{nair2021fixed,athalye2023contractivity}. The novelty of our work lies in the new denoiser we introduce next and in establishing the convergence of PnP-PGD when this denoiser is used.

We now derive the expression for the gradient of the loss function $\ell$ with respect to the inner product defined in~\eqref{eq:innerprod}. Specifically, its gradient $\nabla \! \ell(\X)$ is defined as the unique element in $\bbH$ satisfying
\begin{equation}
\label{eq:defgrad}
\ell(\X + \H) = \ell(\X) + \langle \nabla \! \ell(\X), \H \rangle_{\bbH} + r(\H),
\end{equation}
where the residual $r: \bbH \to \Re$ is such that $r(\H) / \|\H\|_{\bbH} \to 0$ as $\|\H \|_{\bbH} \to 0$. Substituting~\eqref{eq:loss} into~\eqref{eq:defgrad} and performing straightforward calculations, we obtain
\begin{equation*}
\nabla \! \ell (\X) = \A^\top\! \A \X \E \E^\top - \A^\top \Y_h \E^\top + \lambda \X (\E \R)(\E \R)^\top - \lambda \Y_m (\E \R)^\top.
\end{equation*}
Since $\E \E^\top = \I$, this expression can be written more compactly as
\begin{equation}
\label{eq:gradK}
\nabla \! \ell(\X) = \cK(\X) - \A^\top \Y_h \E^\top - \lambda \Y_m (\E \R)^\top,
\end{equation}
where $\cK \in \cL(\bbH)$ is defined as
\begin{equation}
\label{eq:K}
  \cK(\X) = \mathbf{P}_1 \X + \X \mathbf{P}_2,  \qquad  \mathbf{P}_1 := \A^\top \A, \ \mathbf{P}_2 := \lambda (\E \R)(\E \R)^\top.
\end{equation}
In other words, $\nabla \! \ell$ is an affine operator. This will play a crucial role in our subsequent analysis.

\subsection{Kernel Denoiser}
\label{denoiser}

We now describe the denoiser $\cD$ in~\eqref{eq:pnppgd}. The noise in hyperspectral images is generally assumed to be uncorrelated both within and across bands~\cite{aly2014regularized}. However, the spectral bands themselves exhibit strong correlations, and numerous similar patches occur within each band. To exploit the inter-band correlations, we first denoise the image using the kernel-based denoiser proposed in~\cite{nair2019hyperspectral}. To further leverage the intra-band similarity, each band of the denoised image is then processed separately using another kernel denoiser~\cite{sreehari2016plug}.

Specifically, let $\cW : \bbH \to \bbH$ denote the first (high-dimensional) denoiser and $\cV : \bbH \to \bbH$ denote the second (bandwise) denoiser. The overall denoising operator $\cD$ is defined as
\begin{equation}
\label{eq:caskd}
\cD = \cV \circ \cW.
\end{equation}
We refer to $\cW$ as the high-dimensional kernel denoiser, $\cV$ as the bandwise kernel denoiser, and $\cD$ as CasKD (Cascaded Kernel Denoiser).

The mathematical properties of $\cW$ and $\cV$ are central to our convergence analysis. We therefore begin by describing these denoisers in detail. Both $\cW$ and $\cV$ are kernel denoisers~\cite{milanfar}, in which each pixel is denoised by taking a weighted average of its neighboring pixels. The weights are computed using a kernel (or affinity) function. The key difference between $\cW$ and $\cV$ lies in the definition of their respective kernel functions.

We first give the construction of the bandwise denoiser $\cV$, which is more straightforward. For each band $1 \leqslant b \leqslant L_s$, we define a kernel matrix $\K^{(b)}$, whose components are 
\begin{equation}
\label{eq:defKb}
\K^{(b)}_{ij} = h(i-j) \phi_2(\p_i - \p_j)   \qquad (1 \leqslant i,j \leqslant N_m),
\end{equation}
where 
\begin{itemize}
\item $\p_i$ and $\p_j$ are the (vectorized) patches around pixels $i$ and $j$ that are extracted from a guide image, a surrogate of the ground truth constructed from the observed images~\cite{nair2021fixed,gavaskar2020plug}. 
\item $\phi_2$ is a multivariate Gaussian (RBF) kernel,
\[
\phi_2(p_i, p_j) = \exp\left(-\frac{\|p_i - p_j\|_2^2}{2\sigma_2^2}\right),
\]
with $\sigma_2$ denoting the standard deviation and $\|\cdot\|_2$ representing the $\ell_2$ norm on the patch space.

\item $h$ is a symmetric hat function supported on a square window around the origin.
\end{itemize}

By construction, \( \K^{(b)}_{ij} \) takes on large values when pixels $i$ and $j$ and the corresponding patches \( \p_i \) and \( \p_j \) are similar. A simple box function could be used instead of a hat function in~\eqref{eq:defKb}; however, the latter guarantees that the kernel matrix is positive semidefinite~\cite{sreehari2016plug}. 

In the standard construction of a kernel denoiser~\cite{milanfar}, the weight matrix is obtained by normalizing $\K^{(b)}$, 
\begin{equation*}
\D= \diag( \K^{(b)} \,\e),  \quad \W^{(b)}=\D^{-1}\, \K^{(b)}.
\end{equation*}
where $\e$ is the all-ones vector of appropriate length. The problem is that although \(\K^{(b)}\) is symmetric, \(\W^{(b)}\) may not remain symmetric after normalization, the product of two symmetric matrices need not be symmetric. As originally proposed in~\cite{sreehari2016plug}, a symmetric weight matrix can be constructed using the following formula, which we continue to denote by the same symbol:
\begin{equation}
\label{eq:defWb}
\W^{(b)} = \frac{1}{\nu} \D^{-\frac{1}{2}}\K^{(b)} \D^{-\frac{1}{2}} + \diag \Big(\e - \frac{1}{\nu} \, \hat{\e} \Big), 
\end{equation}
where
\begin{equation*}
\D= \diag( \K^{(b)} \,\e), \ \hat{\e} =  \D^{-\frac{1}{2}}\K^{(b)}\D^{-\frac{1}{2}}  \e, \ \nu = \max_i \ \hat{\e}_i.
\end{equation*} 
The bandwise denoiser $\cV: \bbH \to \bbH$ is given by the linear transform
\begin{equation}
\label{eq:defV}
\cV(\X)_b  = \W^{(b)} \X_b \qquad (\X \in \bbH),
\end{equation}
where $\X_b \in \Re^{N_m}$ and $\cV(\X)_b$ denote the $b$-th band of the input and the output of the denoiser. 

The construction of the high-dimensional denoiser $\cW$ is similar but more intricate. Unlike~\eqref{eq:defV}, a single linear transform is applied uniformly across all bands. The main steps in the construction are as follows (see~\cite{nair2019hyperspectral} for more details):

\begin{enumerate}
\item Given the input hyperspectral image represented as a matrix $\X$ of size $N_m  \times L_s$, we extract patches around each pixel while considering all the $L_s$ bands. Specifically, for each pixel $1 \leqslant i \leqslant N_m$, we have a patch vector $\p_i$ of dimension $k^2 L_s$, assuming that a $k \times k$ neighborhood is used around each pixel. 

\item To extract the inter and intra-band correlations among the patches $\p_1,\ldots,\p_{N_s}$, we group them into $C$ clusters with centroids $\boldsymbol{\mu}_1, \dots, \boldsymbol{\mu}_C$.

\item The kernel matrix $\K$ of size $N_m \times N_m$  is defined to be
\begin{equation}
\label{eq:defK}
\K_{ij} = h(i-j) \, \sum_{c=1}^{C} \phi_1(\p_i - \boldsymbol{\mu}_c) \phi_1(\p_j - \boldsymbol{\mu}_c),  \qquad (1 \leqslant i,j \leqslant N_m),
\end{equation}
where \(\phi_1\) is again a Gaussian kernel,
\[
\phi_1(p_i, \mu_c) = \exp\left(-\frac{\|p_i - \mu_c\|_2^2}{2\sigma_1^2}\right),
\]
with \(\sigma_1\) denoting the standard deviation of $\phi_1$.

By construction, \(\K_{ij}\) assumes relatively large values when \(\p_i\) and \(\p_j\) belong to the same cluster.

\item We define a symmetric weight matrix $\W$ of size $N_m \times N_m$ using~\eqref{eq:defWb}, i.e., we set
\begin{equation}
\label{eq:defbW}
\W = \frac{1}{\nu} \D^{-\frac{1}{2}}\K \D^{-\frac{1}{2}} + \diag \Big(\e - \frac{1}{\nu} \, \hat{\e} \Big), 
\end{equation}
where
\begin{equation*}
\D= \diag( \K \,\e), \ \hat{\e} =  \D^{-\frac{1}{2}}\K \D^{-\frac{1}{2}}  \e, \ \nu = \max_i \ \hat{\e}_i.
\end{equation*} 
\item The high-dimensional denoiser $\cW: \bbH \to \bbH$ is given by 
\begin{equation}
\label{eq:defW}
\cW(\X)  = \W \X \qquad (\X \in \bbH).
\end{equation}
In other words, the same weight matrix \( \W \) is applied uniformly across all bands of \( \X \).

\end{enumerate}

The main motivation behind the construction in~\eqref{eq:defK} is that it allows a fast, convolution-based computation of~\eqref{eq:defW}. Further implementation details are provided in~\cite{nair2019hyperspectral}. What is more important, however, is the following observation.

\begin{prop}
\label{prop:WV}
The matrices  $\W^{(b)}$ and $\W$ in~\eqref{eq:defWb} and \eqref{eq:defbW} are (symmetric) positive semidefinite, stochastic, and irreducible.
\end{prop}


Proposition~\ref{prop:WV} is a key result that relies on the intricate construction of the denoisers in Section~\ref{denoiser}. Indeed, starting with a kernel function, constructing a denoiser with the above mathematical properties is a non-trivial task.  Later, we will demonstrate how these properties manifest in the corresponding hyperspectral denoisers \(\cV\) and \(\cW\). 

The technical details behind~Proposition~\ref{prop:WV} can be found in~\cite[Sec.~IV]{sreehari2016plug}. However, one can directly verify from~\eqref{eq:defbW} that $\W$ is symmetric, nonnegative, and $\W\e=\e$ (stochastic). It is also not difficult to show that $\W$ is irreducible; in fact, any operator that performs local averaging on the signal using strictly positive weights is irreducible. The only tricky property is the positive definite property; this can be deduced using the positivity of the Fourier transform of a Gaussian and Bochner's theorem~\cite{sreehari2016plug}.

\begin{figure}[t]
  \centering
  \begin{minipage}{0.19\textwidth}
      \centering
      \includegraphics[width=\linewidth]{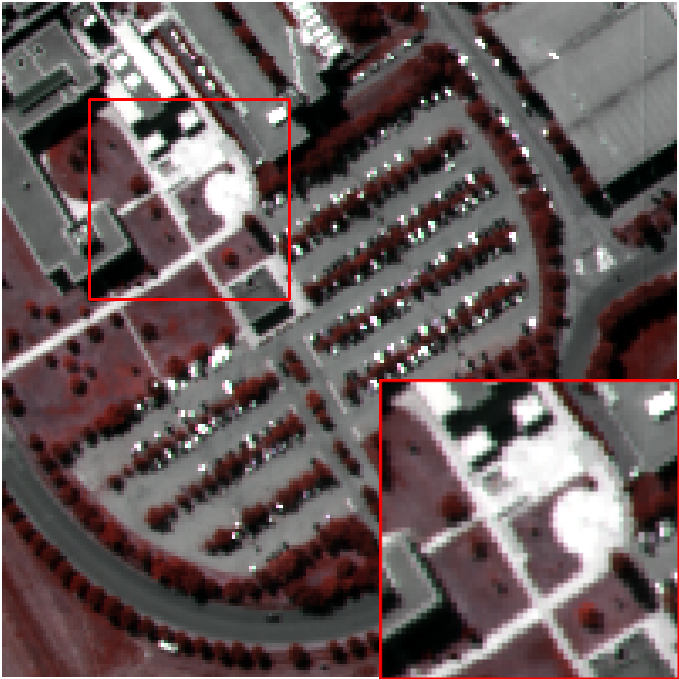}
      \caption*{(a) ground truth}
  \end{minipage} \hfill
  \begin{minipage}{0.19\textwidth}
      \centering
      \includegraphics[width=\linewidth]{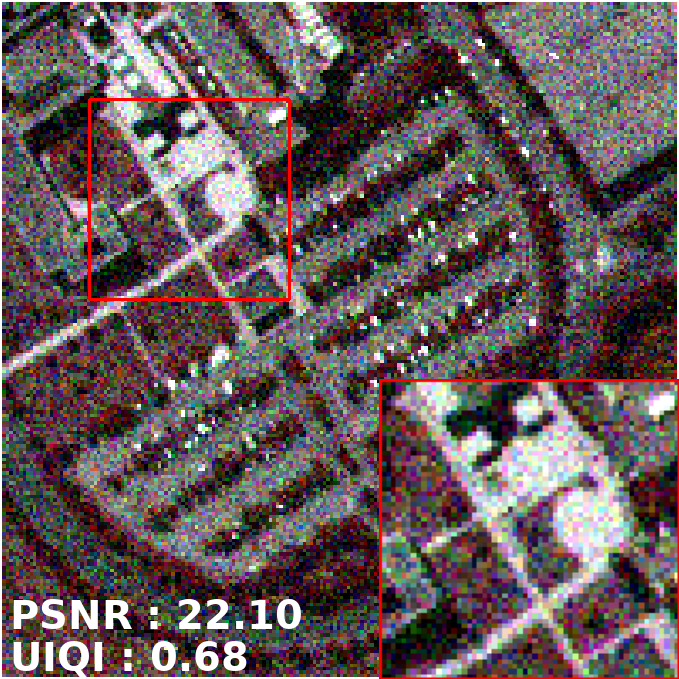}
      \caption*{(b) noisy}
  \end{minipage} \hfill
  \begin{minipage}{0.19\textwidth}
      \centering
      \includegraphics[width=\linewidth]{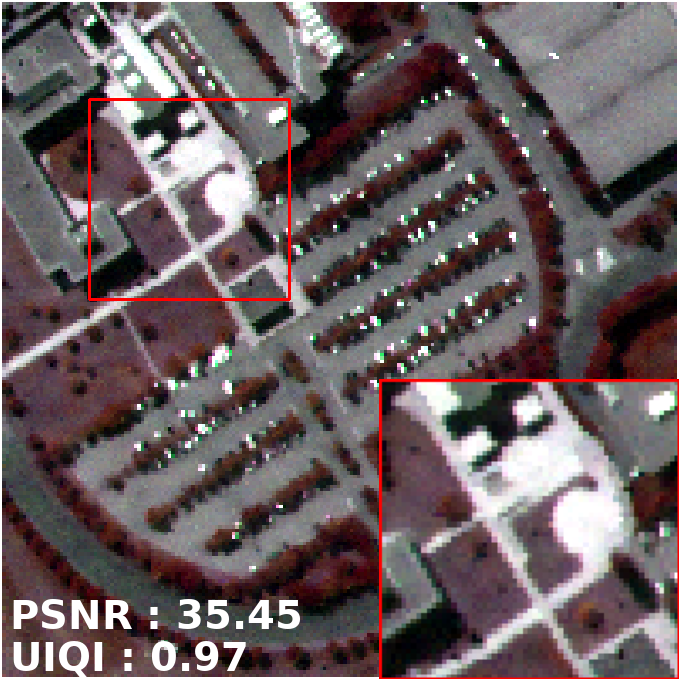}
      \caption*{(c) $\cW$ }
  \end{minipage} \hfill
  \begin{minipage}{0.19\textwidth}
      \centering
      \includegraphics[width=\linewidth]{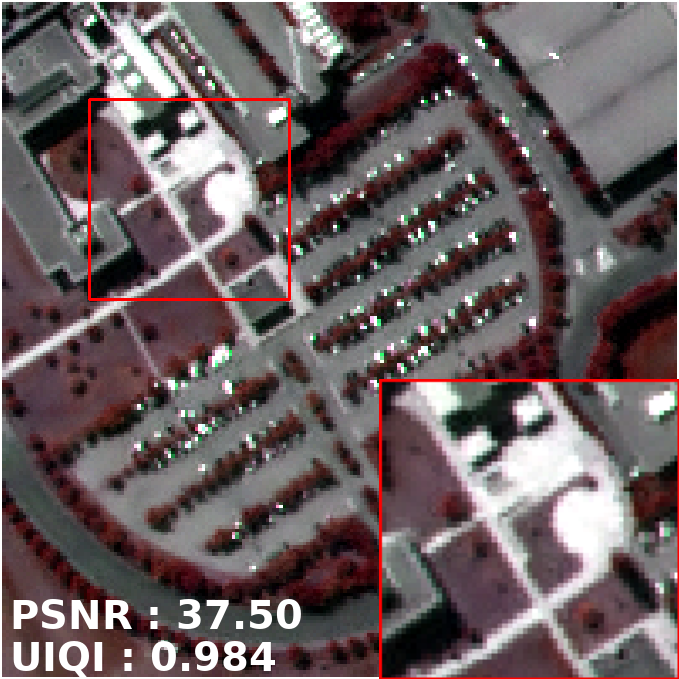}
      \caption*{(d) $\cV$ }
  \end{minipage} \hfill
  \begin{minipage}{0.19\textwidth}
      \centering
      \includegraphics[width=\linewidth]{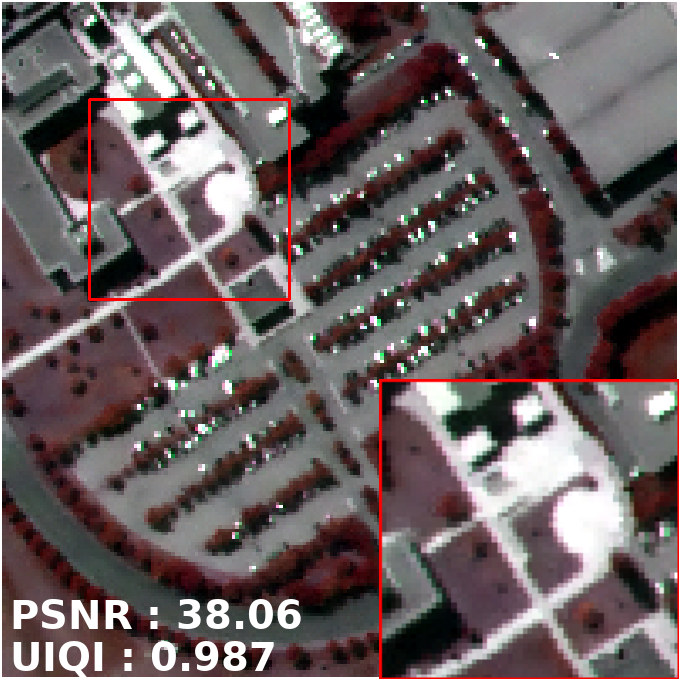}
      \caption*{(e) CasKD }
  \end{minipage}

  \caption{A visual comparison of CasKD on the Pavia dataset and its component denoisers, $\cV$ and $\cW$, in the presence of Gaussian noise with a standard deviation of $20/255$.}
  \label{fig:denoiser_comparison_comp}
\end{figure}

We remark that a different ordering of $\cV$ and $\cW$, such as $\cW \circ \cV$, can also be used in~\eqref{eq:caskd}. In practice, different orderings produce similar performance for both denoising and fusion tasks, and our theoretical results hold for all such combinations. For consistency, we use a single ordering throughout this work. In general, the cascaded versions outperform using $\cV$ or $\cW$ alone, as shown in Table~\ref{denoiser_comp_comparision_gt} and Figure~\ref{fig:denoiser_comparison_comp}.

\begin{table}[t]
\centering
\normalsize
\begin{tabular}{|c| c| c c| c c|}
\hline
{Noise variance} & {Denoiser} & {PSNR} (Pavia) & {UIQI} (Pavia) & {PSNR} (Paris) & {UIQI} (Paris)  \\
\hline
& noisy & 22.10 & 0.68 & 22.10 & 0.68 \\
& $\mathcal{W}$ & 35.45 & 0.973 & 34.52 & 0.961 \\
& $\mathcal{W} \circ \mathcal{W}$ & 36.05 & 0.978 & 34.71 & 0.971 \\
20/255 & $\mathcal{W} \circ \mathcal{V}$ & \textbf{38.01} & \textbf{0.986} & \textbf{36.87} & \underline{$\mathbf{0.982}$} \\
& $\mathcal{V}$ & {37.50} & ${0.984}$ & ${36.80}$ & ${0.981}$ \\
& $\mathcal{V} \circ \mathcal{V}$ & 37.50& 0.984 & 36.80 & 0.981 \\

& $\mathcal{V} \circ \mathcal{W}$ (CasKD) & \underline{$\mathbf{38.06}$} & \underline{$\mathbf{0.987}$} & \underline{$\mathbf{37.06}$} & \underline{$\mathbf{0.982}$} \\
\hline
& noisy & 14.14 & 0.29 & 14.14 & 0.28 \\
& $\mathcal{W}$ & 29.15 & 0.897 & 28.36 & 0.880 \\
& $\mathcal{W} \circ \mathcal{W}$ & 30.24 & 0.920 & 28.75 & 0.887 \\
50/255 & $\mathcal{W} \circ \mathcal{V}$ & \textbf{32.38}& \underline{$\mathbf{0.955}$} & \textbf{30.73} & \textbf{0.915} \\
& $\mathcal{V}$ & $31.53$ & $0.942$ & $30.40$ & $0.922$ \\
& $\mathcal{V} \circ \mathcal{V}$ & 31.53 & 0.942 & 30.40 & 0.922 \\
& $\mathcal{V} \circ \mathcal{W}$ (CasKD) & \underline{$\mathbf{32.43}$} & \underline{$\mathbf{0.955}$} & \underline{$\mathbf{30.86}$} & \underline{$\mathbf{0.931}$} \\
\hline
\end{tabular}
\caption{Comparison of denoising performance on Pavia and Paris. CasKD generally performs better than its individual components, $\cV$ and $\cW$, as well as other combinations, except for $\cW \circ \cV$, for which its performance is comparable. The top two methods are shown in bold, and the best-performing method is underlined.}
\label{denoiser_comp_comparision_gt}
\end{table}

The CasKD module has several hyperparameters that affect the denoising process. Parameters such as the number of clusters and the window size are kept fixed in most experiments (typically \(40\) and \(10 \times 10\)), and the performance is generally robust to moderate changes in these values. However, other parameters---such as the patch size and the denoiser standard deviations \((\sigma_1, \sigma_2)\)---are more problem-dependent. For instance, at higher noise levels, larger values of \((\sigma_1, \sigma_2)\) improve robustness, while increasing the patch size tends to reduce smoothing. Because the denoiser directly affects the reconstruction step in the fusion process, these parameters also influence the final image quality. In our experiments, we set \(\sigma_1\) and \(\sigma_2\) within the range \((5/255, 25/255)\) and fixed the patch size at \(7 \times 7\). We observed that varying these parameters can lead to differences of about 4--5~dB in PSNR and 0.03--0.04 in UIQI, indicating their strong impact on fusion quality. Therefore, selecting appropriate values for \(\sigma_1\), \(\sigma_2\), and the patch size is crucial for achieving optimal performance.

\section{Convergence Analysis}
\label{sec:analysis}

We establish global linear convergence of HyDeFuse in this section. Our analysis builds on the convergence framework of~\cite{athalye2023contractivity}. However, there are key technical differences that prevent a direct application of those results. This will be clarified in \Cref{subsec:discussion}, after we present our results. Proofs of all propositions and lemmas are given in the Appendix.

\subsection{Fixed-Point Analysis}

The starting point of our analysis is to interpret~\eqref{eq:pnppgd} as a fixed-point iteration. Specifically, we can rewrite~\eqref{eq:pnppgd} as
\begin{equation}
\label{eq:fpe}
 \X_{k+1} = \cP(\X_k) + \Q
\end{equation}
where $\cP \in \cL(\bbH)$ and $\Q \in \bbH$ are given by
\begin{equation*} 
 \cP(\X) = \cD \big( \X - \gamma \big( \A^\top\! \A \X  + \lambda \X (\E \R)(\E \R)^\top \big) \big),
\end{equation*} \
and 
\begin{equation*}
\Q = \cD\left( \gamma \A^\top \Y_h \E^\top + \gamma \lambda \Y_m (\E\R)^\top \right).
\end{equation*}
Thus,~\eqref{eq:fpe} can be expressed as 
\begin{equation}
\label{eq:defFP}
\X_{k+1} =\cT(\X_k), \qquad \cT(\X) :=  \cP(\X) + \Q,
\end{equation}
We can further decompose \( \cP \) as 
\begin{equation}
\label{eq:defP}
\cP = \cD\circ \cG,
\end{equation}
where $\cG \in \cL(\bbH)$ is given by
\begin{equation}
\label{eq:defG}
\cG(\X) = \X - \gamma \left( \A^\top\! \A \X + \lambda \X (\E \R)(\E \R)^\top \right).
\end{equation}
We will refer to $\cG$ as the gradient-step operator and the operator $\cT$ as the fixed-point operator. 

As we will see, both \( \cD \) and \( \cG \) are nonexpansive. Hence their composition \( \cP = \cD \circ \cG \), and consequently \( \cT \), is also nonexpansive. However, nonexpansivity alone does not guarantee convergence of the sequence $\{\X_k\}$.  A sufficient condition for convergence is that $\cP$ (and hence $\cT$) is a contraction. In this case, the contraction mapping theorem~\cite{bauschke2011convex} ensures that the sequence $\{\X_k\}$ converges to a unique fixed point for any initialization $\X_0$.

In the rest of this section, we prove that $\cP$ is a contraction operator. Since $\cP \in \cL(\bbH)$, it is a contraction if and only if there exists $\mu \in [0,1)$ such that 
\begin{equation}
\label{eq:defCtrLin}
\|\cP(\X)\|_{\bbH} \leqslant \mu \, \| \X\|_{\bbH} \qquad (\X \in \bbH).
\end{equation}
We have the following observation in this regard.

\begin{prop}
\label{prop:lessthan}
The operator $\cP$ is a contraction if and only if $\|\cP(\X)\| < \|\X\|$ for all nonzero $\X \in \bbH$. 
In this case, we can take
\begin{equation}
\label{eq:kappa}
\mu = \max \big\{ \|\cP(\X)\|_{\bbH} : \ \X \in \bbH, \, \|\X\|_{\bbH} = 1 \big\}
\end{equation}
as the contraction factor.
\end{prop}

One direction is obvious. For the reverse direction, assume that $\|\cP(\X)\| < \|\X\|$ for all nonzero $\X \in \bbH$. 
Note that with $\mu$ defined in \eqref{eq:kappa}, we have \eqref{eq:defCtrLin}; we just have to show that $\mu<1$. Since $\bbH$ is a finite-dimensional vector space, the set $\mathbb{S} =\{ \X \in \bbH: \, \| \X \|_{\bbH} = 1\}$ is compact. Moreover, the map $\X \mapsto \|\cP(\X)\|_{\bbH}$ is continuous on $\mathbb{S}$.
Therefore, by the extreme value theorem, there exists $\X^* \in \mathbb{S}$ such that the maximum in \eqref{eq:kappa} is attained, i.e., $\mu=\| \cP(\X^*) \|_{\bbH}$. It follows that $\mu = \| \cP(\X^*) \|_{\bbH} < \| \X^* \|_{\bbH} = 1$.

\subsection{Constituent Operators}

To show that $\cP$ is a contraction, we need some properties of $\cW, \cV$ and $\cG$. Note that we can express~\eqref{eq:defG} as
\begin{equation}
\label{eq:form}
\cG(\X) = \X - \gamma \cK(\X),
\end{equation}
where $\cK$ is given by~\eqref{eq:K}. We record an important property of $\cK$.

\begin{prop}
\label{prop:betasmooth}
The operator $\cK$ is self-adjoint and $\beta$-Lipschitz with $\beta = \sigma_{\max}(\A)^2 + \lambda \, \sigma_{\max}(\E \R)^2$.
\end{prop}

Since $ \A = \S\B $, it can be shown that $ \sigma_{\max}(\A) \leqslant 1 $. This results in the bound $\beta \leqslant 1 + \lambda \sigma_{\max}(\E\R)^2$,
where $\sigma_{\max}(\E\R)$ can be easily estimated using SVD. A tighter bound can be obtained using the power method~\cite{meyer2023matrix} applied to the linear map $\cK$ in~\eqref{eq:form}. 
Empirical results show that this estimate is often much smaller than the analytical upper bound, although the latter remains useful due to its simplicity and ease of computation.

As $\cK$ is self-adjoint, it follows from~\eqref{eq:form} that $\cG$ is also self-adjoint. Hence, all eigenvalues of $\cG$ are real. We now show that if the step size $\gamma$ in~\eqref{eq:defG} is chosen sufficiently small, then $\cG$ is nonexpansive. Because $\cG$ is self-adjoint, this further implies that $\sigma(\cG) \subset [-1, 1]$. Moreover, we will establish that $-1 \notin \sigma(\cG)$. The result is summarized in the following proposition.

\begin{prop}
\label{prop:G}
For any $0 < \gamma < 2/\beta$, the operator $\cG$ is nonexpansive and $\sigma(\cG)\subset (-1,1]$.
\end{prop}

We note that the nonexpansiveness of the gradient-step operator is a standard result in convex optimization~\cite{bauschke2011convex}. This follows from the Baillon–Haddad theorem~\cite{bauschke2011convex,baillon1977quelques}, a fundamental result for smooth convex functions that applies to our loss function\eqref{eq:loss}. The statement about the spectrum, however, is specific to our operator.

Thus far, we focused on the gradient-step operator $\cG$, one of the two operators in \eqref{eq:defP}. We next analyze the denoiser $\cD$, which is defined in terms $\cW$ and $\cV$.

The properties of $\cW$ and $\cV$ are determined by that of $\W$ and $\W^{(b)}$; see~\eqref{eq:defV} and \eqref{eq:defW}. For instance, $\W$ and $\W^{(b)}$ are symmetric, which forces $\cW$ and $\cV$ to be self-adjoint. The choice of the inner product~\eqref{eq:innerprod} is crucial in this regard. We will also require additional properties of $\cW$ and $\cV$ in our analysis, which are summarized in the following result.

\begin{prop}
\label{prop:VandW}
The operators $\cV$ and $\cW$ are self-adjoint and nonexpansive, $\sigma(\cW) \subset [0,1]$, and $\fix(\cW)$ is a subspace of $\bbH$ consisting of matrices of the form
\begin{equation}
\label{eq:F}
\F=\big[c_1 \e  \ \cdots  \ c_{L_s} \e \big],
\end{equation}
where $c_1, \ldots, c_{L_s} \in \Re$ and $\e \in \Re^{N_m}$ is the all-ones vector.
\end{prop}

Proposition~\ref{prop:VandW} is a key result that stems from the design of $\cW$ and $\cV$ in Section~\ref{denoiser}. Constructing a kernel denoiser with the stated mathematical properties is a nontrivial task. 

We can use the fact that $\cV$ is nonexpansive to simplify the analysis. From~\eqref{eq:caskd}and~\eqref{eq:defP}, we have
\begin{equation}
\label{eq:WG}
\cP = (\cV \circ \cW) \circ \cG = \cV \circ (\cW \circ \cG).
\end{equation}
Hence, if we can show that \(\cW \circ \cG\) is a contraction, it immediately follows that \(\cP\) is also a contraction. In other words, the analysis can be restricted to the two operators $\cW$ and $\cG$.

\subsection{Linear Convergence}

As a first observation, note that if we can find a nonzero $\overline{\X}$ that is a fixed point of both $\cW$ and $\cG$, then $\cW \circ \cG$ cannot be a contraction. Indeed, we would have $(\cW \circ \cG) (\overline{\X})=\overline{\X}$, and consequently 
\begin{equation*}
\| (\cW \circ \cG) (\overline{\X}) \|_{\bbH}= \|\overline{\X}\|_{\bbH},
\end{equation*}
so that $\cW \circ \cG$ cannot be a contraction. In this connection, we have the following result.
\begin{prop}
\label{propfix}
There are no common nonzero fixed points of $\cW$ and $\cG$, namely, $\fix(\cW)  \cap \fix(\cG) =\{\0\}$.
\end{prop}

It turns out that, under suitable assumptions on $\cW$ and $\cG$, the condition $\fix(\cW) \cap \fix(\cG) = {\0}$ is not only necessary but also sufficient to ensure that $\cW \circ \cG$ is a contraction. To establish this, we first state the following result.
\begin{lemma}
\label{lemmaT}
Suppose $\cT \in \cL(\bbH)$ is  self-adjoint and $\sigma(\cT) \subset (-1,1]$. Then for any $\X \in \bbH$, 
\begin{equation*}
\| \cT(\X)\|_{\bbH} = \| \X \|_{\bbH}  \implies \cT(\X) = \X.
\end{equation*}
In particular, if $\X \notin \fix(\cT)$, then $\| \cT(\X)\|_{\bbH} < \| \X \|_{\bbH}$.
\end{lemma}

\vspace{0.3cm}
The above result identifies a condition under which \( \cT\) acts as a contraction in the complement of the subspace \( \fix(\cT) \). This is the closest a linear operator can come to being a contraction if it has a nontrivial $\fix(\cT)$, i.e., if $\fix(\cT) \neq \{\0\}$.  Although a similar conclusion appears in~\cite{athalye2023contractivity}, we include a proof in Appendix~\ref{pf:lemmaT} for completeness. Importantly, Propositions~\ref{prop:G} and~\ref{prop:VandW} confirm that $\cW$ and $\cG$ satisfy the assumptions of Lemma~\ref{lemmaT}.

\vspace{0.3cm}
 We are now ready to establish our main result, namely that $\cT$ is a contraction operator.

\begin{thm}
\label{thm:main}
The fixed-point operator $\cT$ is a contraction. Consequently, for any initialization $\X_0 \in \bbH$, the iterates generated by~\eqref{eq:pnppgd} converge linearly to the unique fixed point of $\cT$.
\end{thm}

\begin{proof}
From~\eqref{eq:WG}, $\cP$, and consequently $\cT$, is a contraction if $\cW \circ \cG$ is a contraction. Furthermore, by Proposition~\ref{prop:lessthan}, it suffices to show that for any $\X \neq \0$,
\begin{equation}
\label{eq:temp}
\| (\cW \circ \cG) (\X)\|_{\bbH} < \|\X\|_{\bbH}.
\end{equation}
There are two possibilities. Suppose $\X \in \fix(\cG)$. In this case, $(\cW \circ \cG )(\X) = \cW  (\X)$. From Proposition~\ref{propfix}, we know that $\X \notin \fix(\cW)$. Since $\cW$ is self-adjoint and $\sigma(\cW) \subset (-1,1]$, Lemma~\ref{lemmaT} implies that $\| \cW  (\X) \|_{\bbH} < \|   \X\|_{\bbH}$. Hence, 
\begin{equation*}
\| (\cW \circ \cG) (\X)\|_{\bbH}  = \| \cW  (\X)\|_{\bbH}  < \|   \X\|_{\bbH}.
\end{equation*}
Consider the other possibility $\X \notin \fix(\cG)$. Here, we have $\| \cG  (\X) \|_{\bbH} < \|   \X\|_{\bbH}$. Thus, since $\cW$ is nonexpansive, 
\begin{equation*}
\| (\cW \circ \cG) (\X)\|_{\bbH}  = \| \cW ( \cG (\X))\|_{\bbH} \leqslant  \| \cG  (\X)\|_{\bbH}  < \|   \X\|_{\bbH}.
\end{equation*}
Thus, we obtain~\eqref{eq:temp} in both cases. It follows that $\cW \circ \cG$, and consequently $\cT$, is a contraction.

By the contraction mapping theorem~\cite{bauschke2011convex}, there exists a unique $\X^* \in \fix(\cT)$. If $\mu \in [0,1)$ denotes the contraction factor of $\cT$, then  
\begin{equation*}
\| \X_{k+1} - \X^* \|_{\bbH} = \| \cT(\X_k) - \cT(\X^*) \|_{\bbH} \leqslant \mu \| \X_k - \X^* \|_{\bbH}.
\end{equation*}
By iterating this inequality, we obtain for all $k \geqslant 0$,
\begin{equation}
\label{eq:cvgrate}
\| \X_k - \X^* \|_{\bbH}  \leqslant \mu^k \| \X_0 - \X^* \|_{\bbH},
\end{equation}
which establishes the global linear convergence of the sequence $\{\X_k\}$. 
\end{proof}

From~\eqref{eq:cvgrate}, it follows that the convergence rate is directly governed by the contraction factor~$\mu$. A smaller $\mu$ yields faster convergence of the iterates, whereas values of $\mu$ closer to~$1$ lead to slower, yet still guaranteed, convergence.


\subsection{Discussion}
\label{subsec:discussion}

We now discuss how our analysis and results connect with earlier works that employ kernel denoisers to construct provably convergent PnP models~\cite{nair2019hyperspectral,sreehari2016plug,gavaskar2021plug,athalye2023contractivity}. These studies interpret the kernel denoiser as the proximal operator of a convex regularizer, thereby linking the PnP framework to a convex optimization setting. The convergence of the corresponding objective function is then established using standard results from proximal algorithm theory~\cite{bauschke2011convex}.

In contrast, the CasKD denoiser in HyDeFuse is not necessarily self-adjoint, since $\cV$ and $\cW$ do not commute. This asymmetry implies that $\cD$ cannot be expressed as the proximal operator of any convex regularizer, this follows from a classical result due to Moreau~\cite{Moreau1965,chaudhury:hal-05038838}. Consequently, the approach in~\cite{nair2019hyperspectral,sreehari2016plug,gavaskar2021plug} does not directly apply here.

Moreover, we establish a stronger result, namely the convergence of the iterates themselves (not just the objective), and further provide a linear convergence rate. While our approach is inspired by~\cite{athalye2023contractivity}, the specific loss function, vector space, operators, denoiser, and supporting results differ substantially in our setting. Therefore, the analysis in~\cite{athalye2023contractivity} cannot be directly extended to HyDeFuse.

\section{Numerical Results}

\label{sec:results}

We validate our theoretical results and demonstrate HyDeFuse in practice. We present numerical results on global convergence and convergence rate. We compare the proposed CasKD denoiser with existing ones and benchmark HyDeFuse against both classical and state-of-the-art fusion methods.

\subsection{Experimental Setup}
\label{subsec:para}

We begin by describing the dataset, the parameters of the forward model, the quality metrics used for evaluation, and the denoising and fusion methods used for comparison.

\textbf{Datasets}. We use standard datasets for our experiments: Pavia~\cite{simoes2014convex}, Paris~\cite{simoes2014convex}, and Chikusei~\cite{nair2019hyperspectral}. The Pavia dataset, captured by the ROSIS sensor, contains 115 s,pectral bands with a spectral range of (0.43-0.56)~$\mu$m. We cropped the image to a size of $200 \times 200 \times 93$ by removing bands with very low SNR~\cite{simoes2014convex}. The Paris dataset was captured by two Earth Observation-1 (EO-1) satellite sensors: the Hyperion instrument and the Advanced Land Imager (ALI). The original image has a size of $72 \times 72 \times 128$. The Chikusei dataset, captured by Headwall’s Hyperspec-VNIR-C imaging sensor, contains 128 spectral bands with a range of (0.363-1.018)~$\mu$m. The Chikusei image used in our experiments has a size of $540 \times 480 \times 128$.

\textbf{Model parameters.}  To test robustness under different conditions, we model the blurring operator for the  HS image in two ways. The main setup uses the Starck–Murtagh filter, as in~\cite{simoes2014convex,nair2019hyperspectral}. Following~\cite{dian2024hyperspectral}, we consider a Gaussian point spread function (PSF) with a radius of 7 and a standard deviation of 2. The downsampling factors used to create the observed HS image are 4 for Pavia, 6 for Chikusei, and 3 for Paris. The spectral response matrix $\R$ is computed as described in~\cite{simoes2014convex}. Unless stated otherwise, the signal-to-noise ratio (SNR) is fixed at 20~dB for both the HS and MS images, denoted by $\mathrm{SNR}_h$ and $\mathrm{SNR}_m$. The HS images of Pavia, Chikusei, and Paris contain $L_h = 93$, $128$, and $128$ bands, while the corresponding MS images have $L_m = 4$, $8$, and $9$ bands.

\textbf{Hyperparameters}. Following~\cite{simoes2014convex}, we assume that the ground-truth image lies in a low-dimensional subspace, with dimension $L_s = 10$ for the Pavia and Paris datasets and $L_s = 5$ for the Chikusei dataset. The patch size is set to $7 \times 7$, the search window to $10 \times 10$, and the number of clusters to $40$. The parameters $\sigma_1$ and $\sigma_2$ are chosen from the range $(5/255, 25/255)$ and tuned for each dataset. In our experiments, we fix the number of iterations to 100 rather than using an explicit stopping rule. However, a stopping condition, such as $\|\mathbf{X}_k - \mathbf{X}_{k-1}\|_{\bbH}/\|\mathbf{X}_k\|_{\bbH} < \varepsilon$, can be easily integrated into our framework. The theoretical convergence of the iterative scheme guarantees that this condition will always be met for any chosen tolerance $\varepsilon$.

\textbf{Compared methods}. We compare our CasKD denoiser with both traditional and recent denoisiners. Specifically, we benchmark CasKD against BM4D~\cite{maggioni2012nonlocal}, DnCNN~\cite{zhang2017beyond}, FHDD-NLM~\cite{nair2019hyperspectral}, KBR~\cite{kbr}, LLRT~\cite{LLRT}, FastHyMix~\cite{zhuang2021fasthymix}, and NGMeet~\cite{NGMeet}. For image fusion, we evaluate HyDeFuse against a range of classical and state-of-the-art methods, including GSA~\cite{aiazzi2007improving}, CNMF~\cite{yokoya2011coupled}, GLP~\cite{aiazzi2006mtf}, MAPSMM~\cite{eismann2004resolution}, HySure~\cite{simoes2014convex}, GTTN~\cite{dian2024hyperspectral}, and CTDF~\cite{xu2024coupled}.

\textbf{Quality metrics}. We employ standard quality metrics~\cite{simoes2014convex,dian2020regularizing,dian2024hyperspectral}, including Peak Signal-to-Noise Ratio (PSNR), Root Mean Squared Error (RMSE), Spectral Angle Mapper (SAM), Erreur Relative Globale Adimensionnelle de Synthèse (ERGAS), and Universal Image Quality Index (UIQI). For reconstruction tasks, lower values of RMSE, SAM, and ERGAS indicate better performance, whereas higher PSNR and UIQI values (with a maximum of $1$) correspond to superior reconstruction quality.

\subsection{Denoising Results}

The CasKD denoiser combines two kernel denoisers. While combining denoisers need not always improve performance,  the two components play different roles in CasKD—they handle inter- and intra-band correlations between patches. This likely explains the improved results seen in Table~\ref{denoiser_comp_comparision_gt}.

\begin{figure}[H]
  \centering
  \begin{minipage}{0.32\textwidth}
      \centering
      \includegraphics[width=\linewidth]{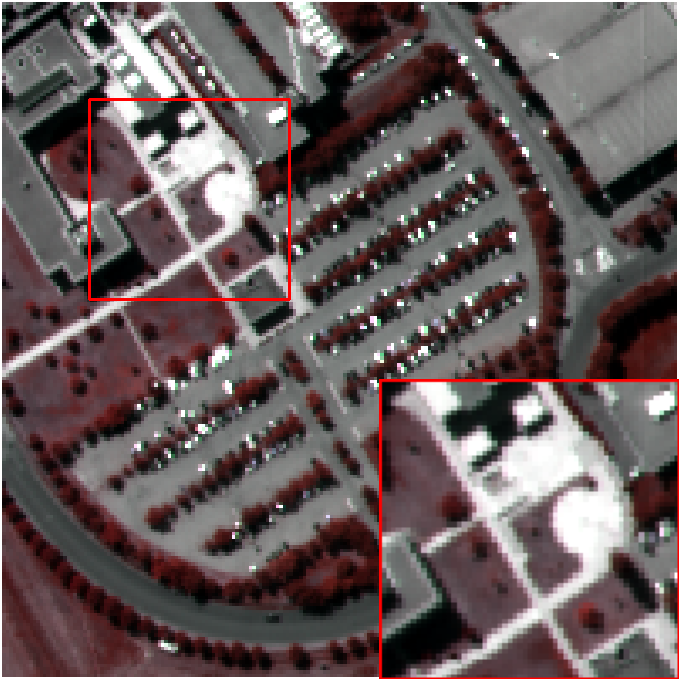}
      \caption*{{(a) ground truth}}
  \end{minipage} \hfill
  \begin{minipage}{0.32\textwidth}
      \centering
      \includegraphics[width=\linewidth]{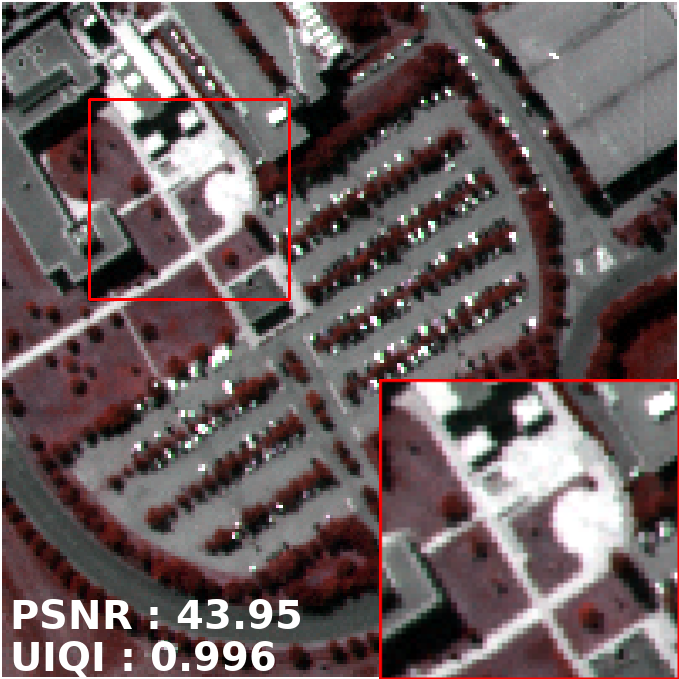}
      \caption*{{(b) FHDD-NLM} }
  \end{minipage} \hfill
  \begin{minipage}{0.32\textwidth}
      \centering
      \includegraphics[width=\linewidth]{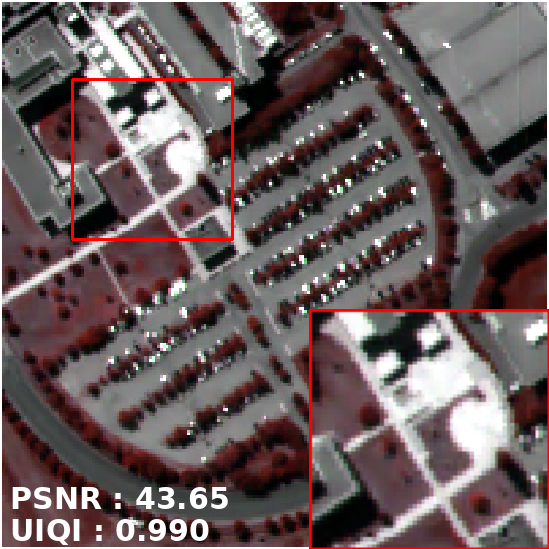}
      \caption*{{(c) BM4D} }
  \end{minipage}

  \vspace{.01cm} 

  \begin{minipage}{0.32\textwidth}
      \centering
      \includegraphics[width=\linewidth]{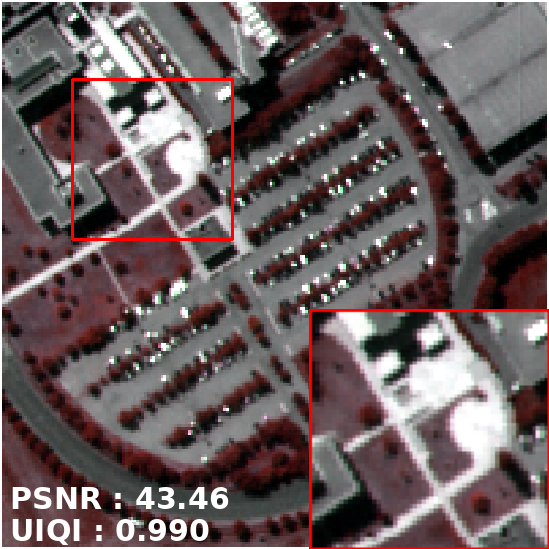}
      \caption*{{(d) DnCNN}}
  \end{minipage} \hfill
  \begin{minipage}{0.32\textwidth}
      \centering
      \includegraphics[width=\linewidth]{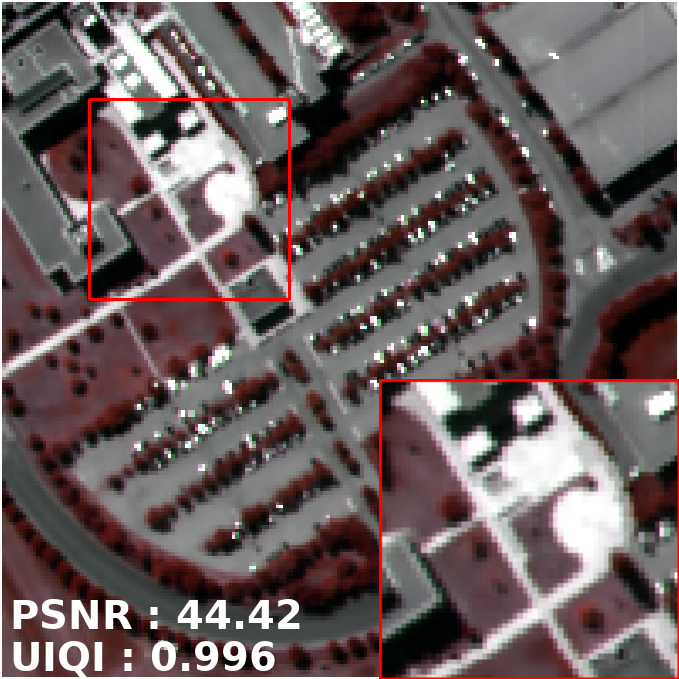}
      \caption*{{(e) KBR} }
  \end{minipage} \hfill
  \begin{minipage}{0.32\textwidth}
      \centering
      \includegraphics[width=\linewidth]{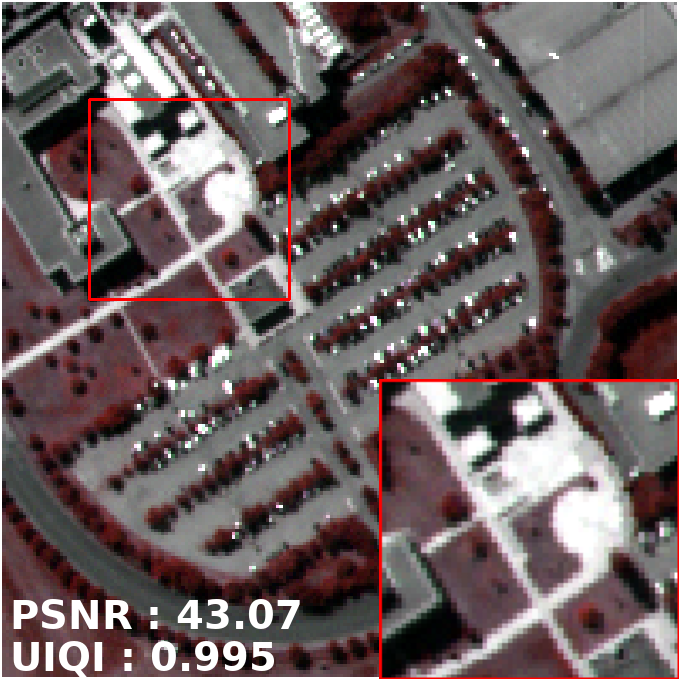}
      \caption*{{(f) LLRT} }
  \end{minipage}

  \vspace{.01cm} 

  \begin{minipage}{0.32\textwidth}
      \centering
      \includegraphics[width=\linewidth]{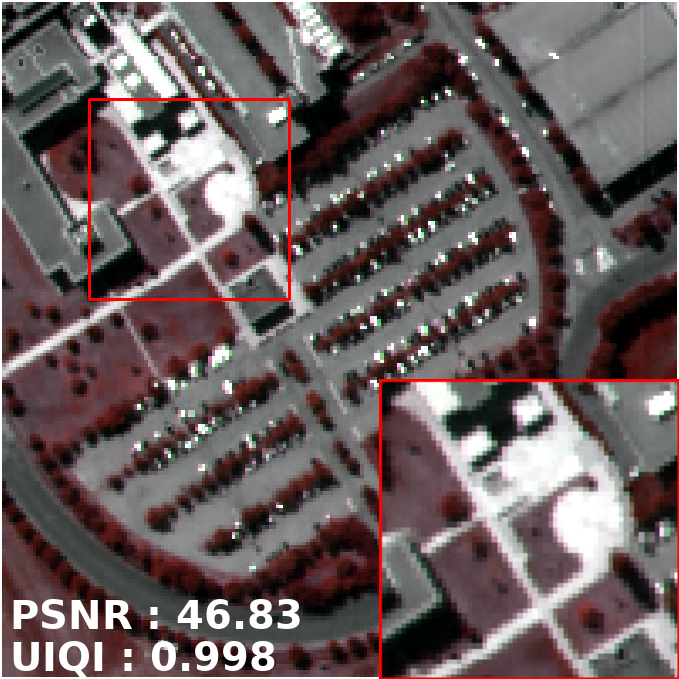}
      \caption*{{(g) FastHyMix} }
  \end{minipage} \hfill
  \begin{minipage}{0.32\textwidth}
      \centering
      \includegraphics[width=\linewidth]{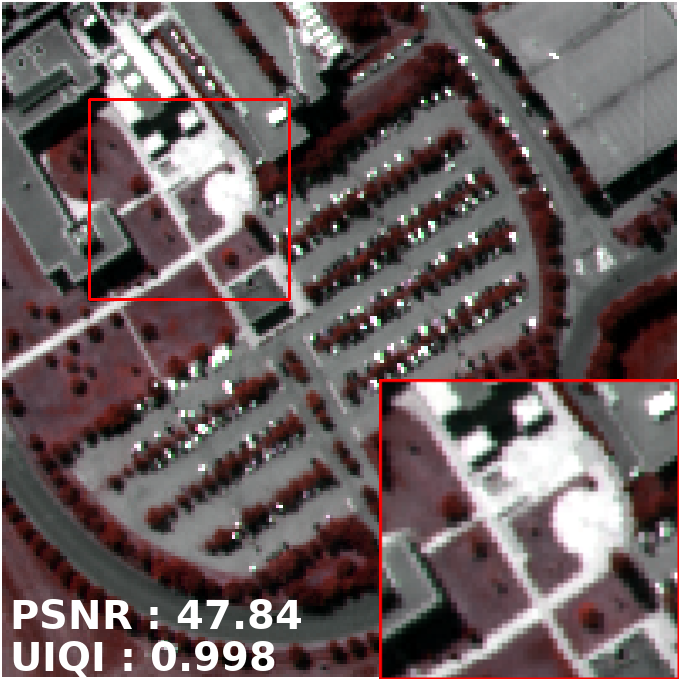}
      \caption*{{(h) NGMeet} }
  \end{minipage} \hfill
  \begin{minipage}{0.32\textwidth}
      \centering
      \includegraphics[width=\linewidth]{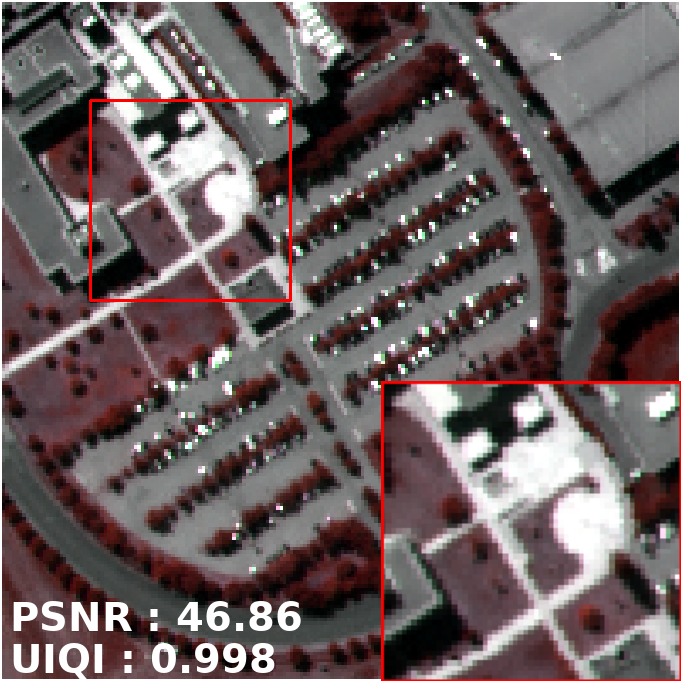}
      \caption*{{(i) CasKD} }
  \end{minipage}

\caption{Visual comparison of CasKD with other denoisers on the Pavia dataset. Zero-mean Gaussian noise with a standard deviation of $5/255$ is independently added to pixels across all bands. For visualization, three selected bands are shown.}
  \label{fig:den_comp2}
\end{figure}

To further assess CasKD, we compare it with several well-known denoisers: BM4D~\cite{maggioni2012nonlocal}, DnCNN~\cite{zhang2017beyond}, FHDD-NLM~\cite{nair2019hyperspectral}, KBR~\cite{kbr}, LLRT~\cite{LLRT}, FastHyMix~\cite{zhuang2021fasthymix}, and NGMeet~\cite{NGMeet}. A visual comparison is presented in Figure~\ref{fig:den_comp2}. A small region is marked with a red box, and its zoomed-in view is shown in the lower-right corner to help access the denoising quality. Because the noise is relatively low ($\sigma = 5/255$), most denoisers produce visually good results; however, the insets make it easier to observe subtle texture differences among the methods.

Table~\ref{tab:den_comp3} provides a detailed quantitative comparison using PSNR and UIQI. CasKD performs competitively with existing denoisers and, in some cases, achieves the best results.

\begin{table}[ht]
\centering
\normalsize
\begin{tabular}{|c|c c|c c|}
\hline

Methods & PSNR (5/255) & UIQI (5/255) & PSNR (20/255) & UIQI (5/255) \\
\hline
Noisy & 33.20 & 0.95 & 22.10 & 0.68 \\
FHDD-NLM & 43.22 & 0.985 & 35.45 & 0.97 \\
BM4D & 43.65 & 0.989 & 35.01 & 0.97 \\
DnCNN & 43.46 & 0.990 & 37.83 & 0.98 \\
KBR & 44.42 & 0.996 & 37.75 & 0.98 \\
LLRT & 43.07 & 0.994 & $\mathbf{38.92}$ & 0.97 \\
FastHyMix & 46.83 & \underline{$\mathbf{0.998}$} & 38.76 & 0.980 \\
NGMeet & \underline{$\mathbf{47.84}$} & \underline{$\mathbf{0.998}$} & \underline{$\mathbf{40.83}$} & $\mathbf{0.983}$ \\
CasKD & $\mathbf{46.86}$ & \underline{$\mathbf{0.998}$} & 38.06 & \underline{$\mathbf{0.987}$} \\
\hline
\end{tabular}
\caption{Comparison of denoising results (with Gaussian noise) on the Pavia dataset for zero-mean Gaussian noise with standard deviations of 5/255 and 20/255 (shown in brackets). The top two values are in bold, and the best value is underlined. This convention is used in all the tables.}
\label{tab:den_comp3}
\end{table}

\subsection{Convergence Results}

We evaluate the convergence behavior of HyDeFuse under various settings. First, we analyze the empirical convergence rate and present evidence of its contractive nature. Next, we examine the effect of the step size on convergence and its relation to linear convergence. We then assess the stability of the reconstruction in terms of PSNR for the same experiments. Finally, we test the uniqueness of the fixed point of HyDeFuse.

We first estimate the contraction factor $\mu$ in~\eqref{eq:defCtrLin}, which, as shown in~\eqref{eq:cvgrate}, determines the convergence rate of HyDeFuse. We use the power method~\cite{meyer2023matrix} to compute $\mu$, the largest singular value of the linear operator $\cP$. A direct computation using SVD is impractical, since $\cP$ is too large to store explicitly. Instead, the power method approximates $\mu$ by iteratively applying $\mathcal{P}$ and its adjoint, which is given by $\mathcal{P}^* = \mathcal{G} \circ \mathcal{W} \circ \mathcal{V}$, since $\mathcal{V}$, $\mathcal{W}$, and $\mathcal{G}$ are self-adjoint.

\begin{figure}[t]
  \centering
  \begin{minipage}{0.9\linewidth}
    \centering
    \includegraphics[width=\linewidth]{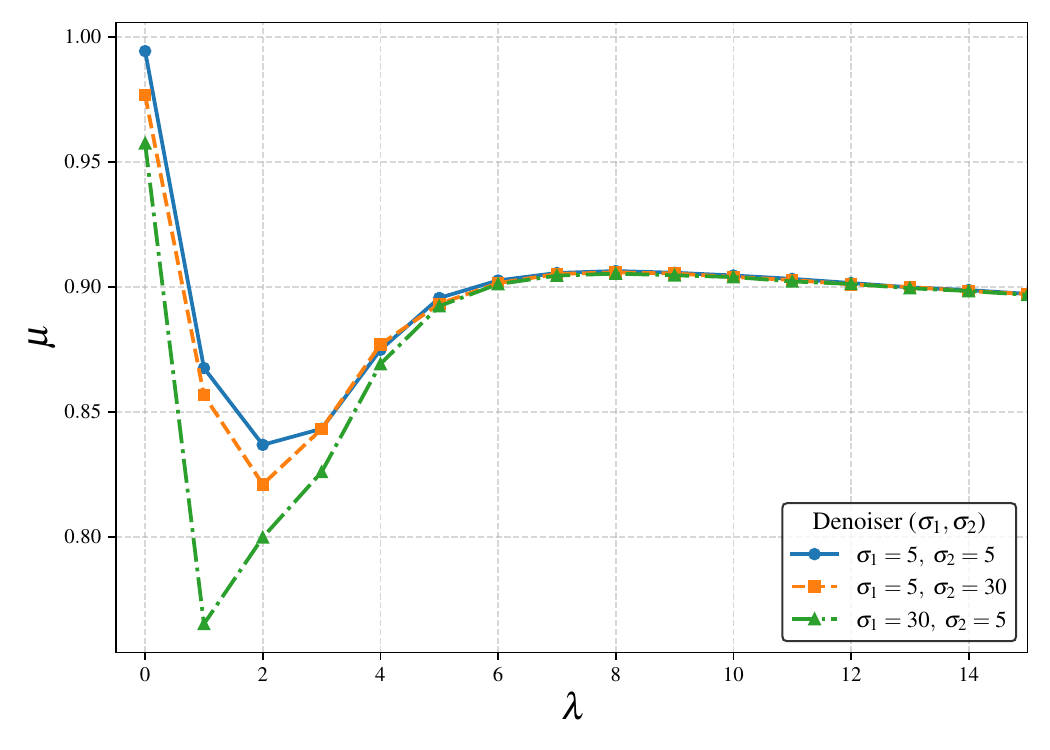}
  \end{minipage}
  \caption{Variation of the convergence rate $\mu$ for different values of $\lambda$ in the loss function~\eqref{eq:loss}, evaluated on the Chikusei dataset. The parameters $\sigma_1$ and $\sigma_2$ (reported without scaling by $1/255$) represent the standard deviations of the Gaussian (RBF) kernels used in $\mathcal{W}$ and $\mathcal{V}$, respectively.}
  \label{fig:contra}
\end{figure}

\begin{table}[H]
\centering
\normalsize
\begin{tabular}{|c|c|c|c|c|c|c|}
\hline

$\gamma$ & $(a,a)$ & $(b,a)$ & $(c,a)$ & $(a,b)$ & $(b,b)$ & $(c,c)$ \\
\hline
0.1 & 0.985 & 0.986 & 0.986 & 0.986 & 0.986 & 0.986 \\
0.5 & 0.961 & 0.964 & 0.964 & 0.964 & 0.964 & 0.964 \\
1 & 0.922 & 0.929 & 0.929 & 0.929 & 0.929 & 0.929 \\
1.5 & 0.884 & 0.894 & 0.894 & 0.894 & 0.894 & 0.894 \\
1.8 & 0.861 & 0.873 & 0.873 & 0.873 & 0.873 & 0.873 \\
1.9 & 0.906 & 0.897 & 0.897 & 0.897 & 0.897 & 0.896 \\
\hline
\end{tabular}
\caption{Contraction factors are shown for different step sizes and denoiser settings. These results are evaluated on the Chikusei dataset. The step size $\gamma$ is expressed in multiples of $1/\beta$, where $\beta$, computed using the power method, is found to be $0.9757$. The pair $(\sigma_1, \sigma_2)$ represents the standard deviations of the Gaussian (RBF) kernels used in $\mathcal{W}$ and $\mathcal{V}$, with $a = 5/255$, $b = 10/255$, and $c = 30/255$.}
\label{tab:contra1}
\end{table}

In Table~\ref{tab:contra1}, we validate the contraction factor $\mu $ for HyDeFuse using the Chikusei dataset. The value of $\mu$ depends on the step size $\gamma$, the forward model parameters $\A$ and $\R$, and the denoiser $\cD$. We compute $\mu$ using the power method for step sizes $0 < \gamma < 2/\beta$ and for different denoiser settings. As predicted by Theorem~\ref{thm:main}, the contraction factor remains below $1$.

We also compute $\mu$ for different values of $\lambda$, which controls the balance in the loss function. The results, shown in Figure~\ref{fig:contra}, again confirm that $\mu < 1$. As seen in the plot, $\mu$ approaches $1$ as $\lambda$ tends to $0$, but remains slightly below it, with $\mu \approx 1 - \varepsilon$ where $\varepsilon \approx 10^{-3}$. Since these values are very close to $1$, a  mathematical proof is important to rule out numerical artifacts that could make $\mu$ appear smaller than $1$.

Additionally, we see that convergence is faster when $\lambda$ is neither too large nor too small, which is noteworthy as this range is also expected to yield optimal fusion performance. For the Chikusei dataset (Figure~\ref{fig:contra}), optimal fusion performance is observed for $\lambda$ in the range 4–8. This is consistent with the observation that intermediate $\lambda$ values yield optimal convergence. A similar trend is seen in the Pavia and Paris datasets.
\begin{figure}[t]
  \centering
   \includegraphics[width=1\linewidth]{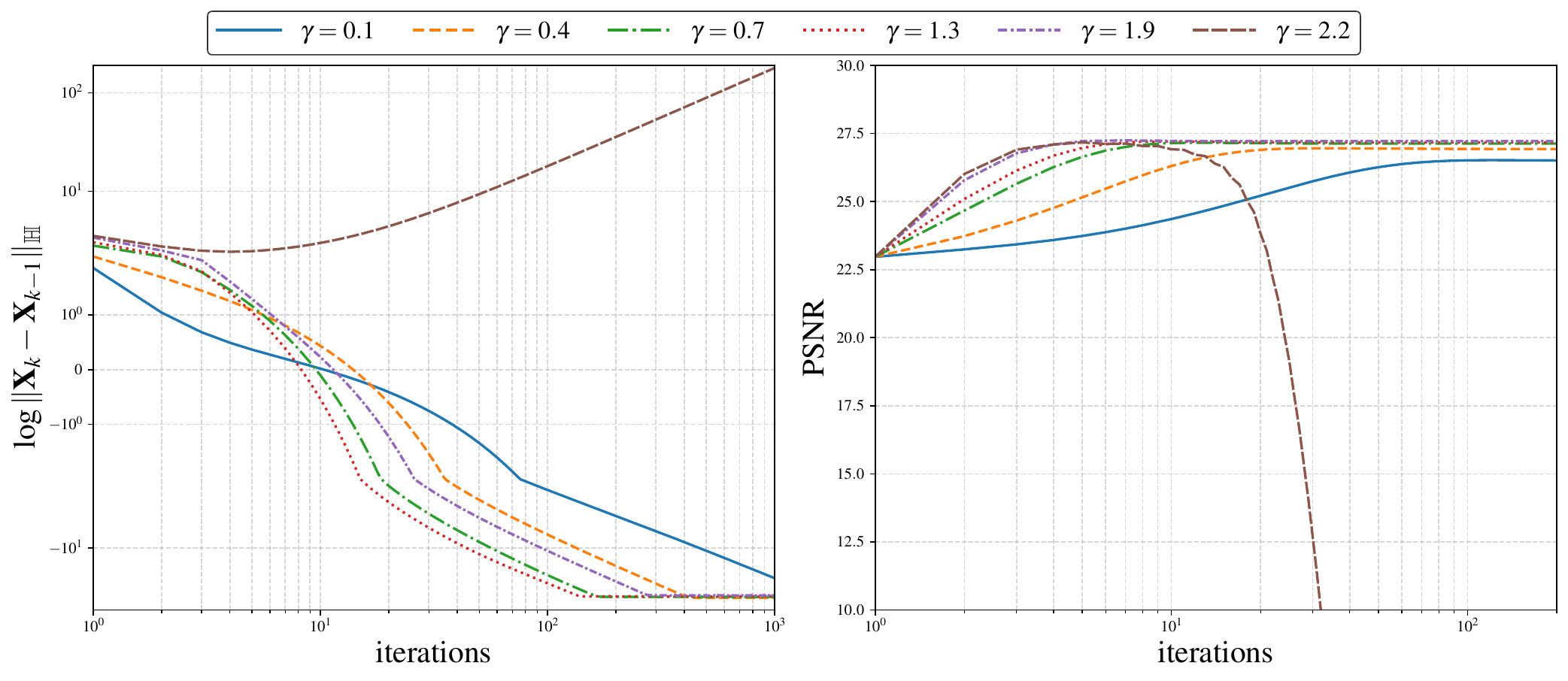}
    \caption{The plot shows the successive differences between HyDeFuse iterates and their corresponding PSNR values for different step sizes $\gamma$ (in units of $1/\beta$) on the Paris dataset. Using the power method, we find $\beta = 0.5998$ for this setup. Convergence is achieved for $\gamma \in (0, 2/\beta)$, but as seen with $\gamma = 2.2/\beta$, it fails outside this range. The PSNR stabilizes within about 10 iterations for an optimal choice of $\gamma$.}
  \label{fig:linear_conv}
\end{figure}

\begin{figure}[t]
  \centering
  \begin{minipage}{0.24\textwidth}
      \centering
      \includegraphics[width=\linewidth]{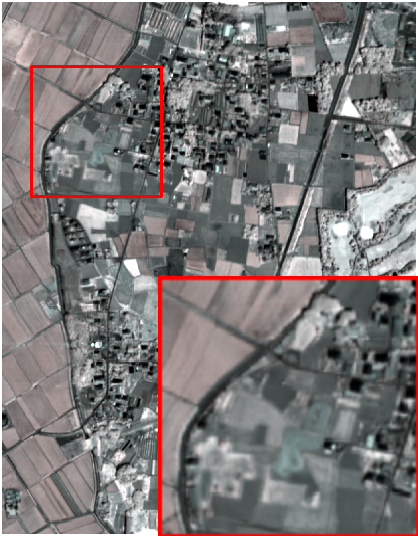} 
      \caption*{{(a) ground truth}} 
     
  \end{minipage} \hfill
  \begin{minipage}{0.24\textwidth}
      \centering
      \includegraphics[width=\linewidth]{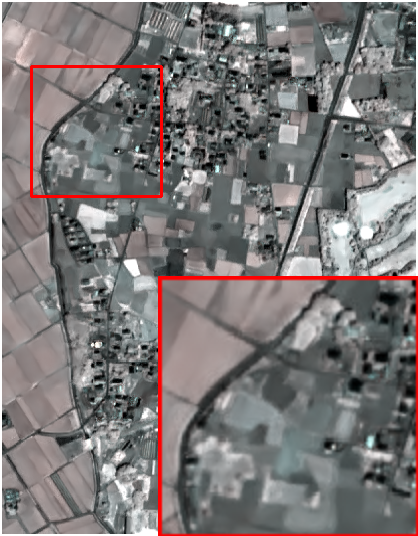} 
      \caption*{{(b) $\X_0$ : all-ones}} 
      
  \end{minipage} \hfill
  \begin{minipage}{0.24\textwidth}
      \centering
      \includegraphics[width=\linewidth]{output_chikuksie_pgd_same_intialization.png} 
      \caption*{{(c) $\X_0$ : all-zeros }} 
      
  \end{minipage} \hfill
  \begin{minipage}{0.24\textwidth}
      \centering
      \includegraphics[width=\linewidth]{output_chikuksie_pgd_same_intialization.png} 
      \caption*{{(d) $\X_0$ : white noise}} 
      
  \end{minipage}
  \caption{HyDeFuse results on the Chikusei dataset for different initializations $\mathbf{X}_0$. In all cases, the performance metrics remain identical (PSNR = 42.27, RMSE = 0.0076, ERGAS = 1.39, SAM = 1.79, and UIQI = 0.967), confirming that the HyDeFuse iterates converge to the same fixed point.}
  \label{fig:fix}
\end{figure}

We next study the convergence of the iterates $\{\mathbf{X}_k\}$ generated using~\eqref{eq:pnppgd}. Figure~\ref{fig:linear_conv} shows a sample result on the Paris dataset using the Starck–Murtagh filter as the blurring operator. As predicted by Theorem~\ref{thm:main}, the successive difference $\|\X_k - \X_{k-1}\|_\bbH$ decreases to zero, and the PSNR stabilizes within about 10–15 iterations a trend consistently observed across all datasets.

When the step size $\gamma$ exceeds the admissible range $0<\gamma<2/\beta$, we see in Figure~\ref{fig:linear_conv} that the algorithm diverges, with $\|\X_k - \X_{k-1}\|_\bbH$ gradually increasing. This shows the importance of choosing $\gamma$ properly. Moreover, we notice that increasing $\gamma$ does not necessarily speed up convergence.

To further demonstrate the contractive nature of the reconstruction operator $\mathcal{T}$, we perform fusion experiments on the Chikusei dataset with different initializations $\mathbf{X}_0$. As shown in~\Cref{fig:fix}, all runs converge to the same reconstruction, confirming the uniqueness of the fixed point of the contractive operator $\mathcal{T}$.

The results in this subsection provide clear experimental evidence of the convergence of HyDeFuse. Using the power method, we show that the reconstruction operator $\mathcal{P}$ is contractive across different settings, with the contraction factor $\mu$ always less than $1$. This holds for various step sizes $\gamma$ and values of the balancing parameter $\lambda$, consistent with Theorem~\ref{thm:main}. The iterates exhibit linear convergence, and the PSNR stabilizes within 10–15 iterations. We also find that large values of $\gamma$ can lead to divergence, emphasizing the need for proper parameter selection. Finally, the fixed-point analysis confirms that HyDeFuse consistently converges to a unique solution from different initializations.

\subsection{Fusion Results}
Finally, we compare the performance of HyDeFuse with several classical methods as well as recent state-of-the-art techniques, under two different noise levels.

\begin{figure}[t]
  \centering
  \begin{minipage}{0.24\textwidth}
      \centering
      \includegraphics[width=\linewidth]{output_pavaigrndlatest.png}
      \caption*{{(a) ground truth}} 
  \end{minipage} \hfill
  \begin{minipage}{0.24\textwidth}
      \centering
      \includegraphics[width=\linewidth]{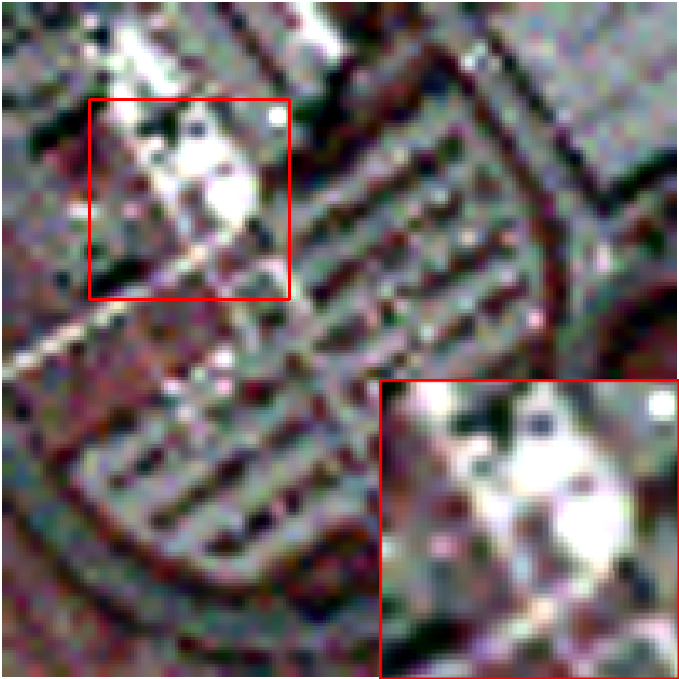}
      \caption*{{(b) Bicubic}}
  \end{minipage} \hfill
  \begin{minipage}{0.24\textwidth}
      \centering
      \includegraphics[width=\linewidth]{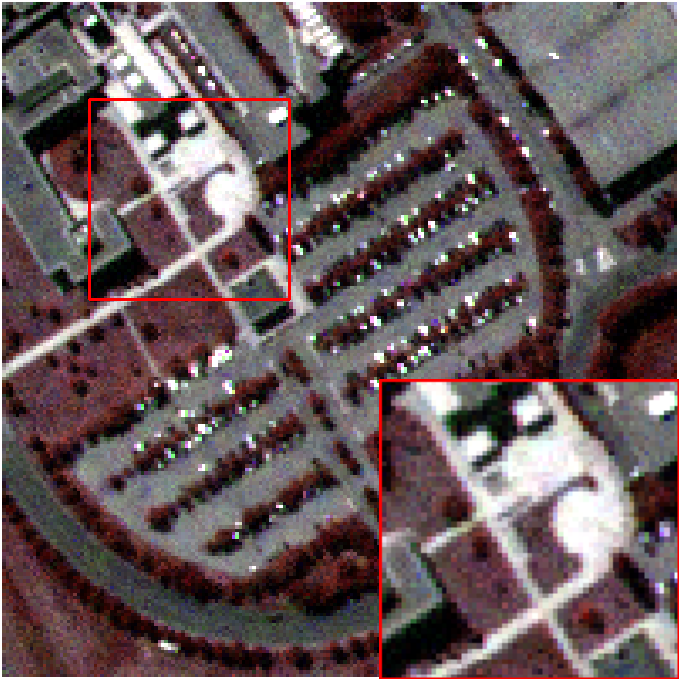}
      \caption*{{(c) CNMF}}
  \end{minipage} \hfill
  \begin{minipage}{0.24\textwidth}
      \centering
      \includegraphics[width=\linewidth]{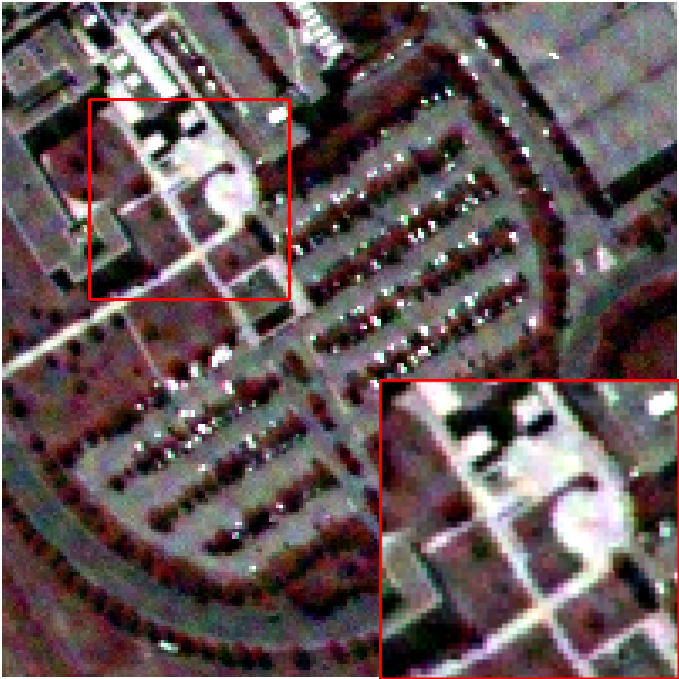} 
      \caption*{{(d) GLP}} 
  \end{minipage}

  \vspace{0.5cm} 
  
   \begin{minipage}{0.24\textwidth}
      \centering
      \includegraphics[width=\linewidth]{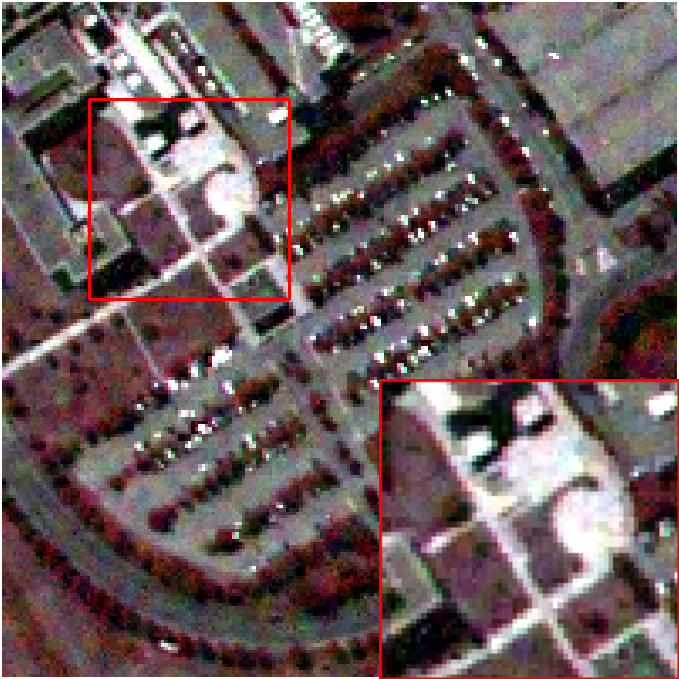} 
      \caption*{{(e) MASMM}} 
  \end{minipage}\hfill
   \begin{minipage}{0.24\textwidth}
      \centering
      \includegraphics[width=\linewidth]{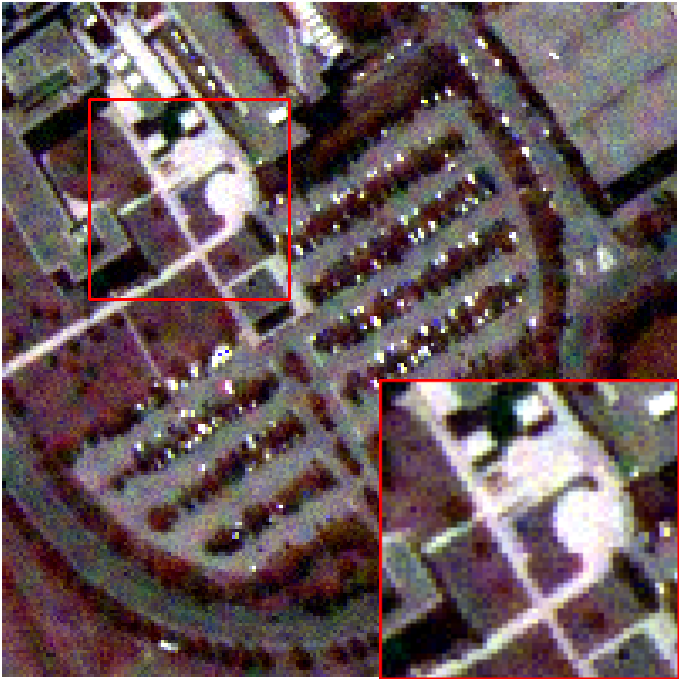} 
      \caption*{{(f) GSA}} 
  \end{minipage}\hfill
 \begin{minipage}{0.24\textwidth}
      \centering
      \includegraphics[width=\linewidth]{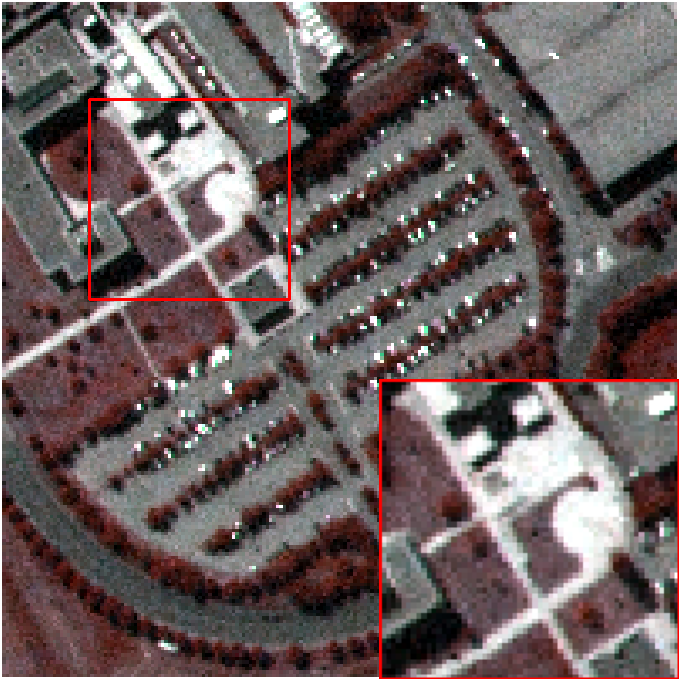} 
      \caption*{{(g) Hysure}} 
  \end{minipage}\hfill
  \begin{minipage}{0.24\textwidth}
      \centering
      \includegraphics[width=\linewidth]{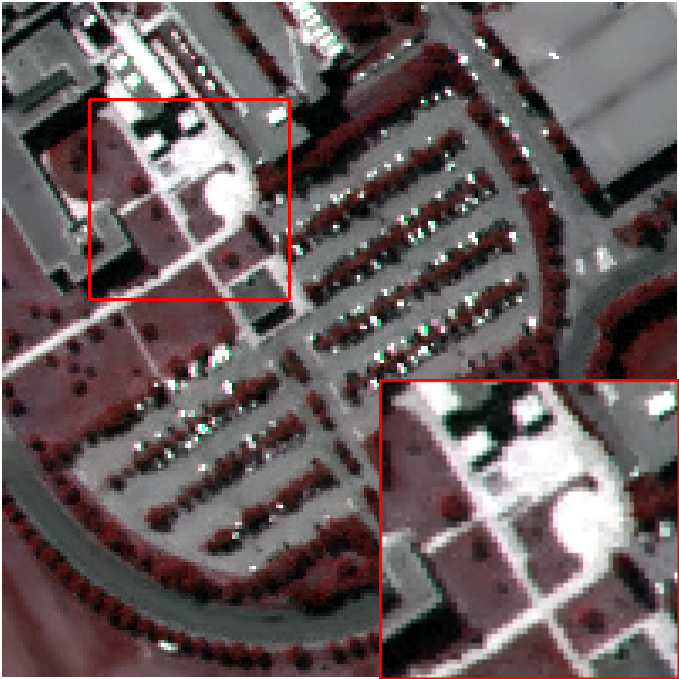} 
      \caption*{{(h) HyDeFuse}} 
  \end{minipage}

  \caption{Fusion results on the Pavia dataset. The quality metrics are shown in Table \ref{tab:fused_old}.}
  \label{fig:fused1}
\end{figure}

A comparison between HyDeFuse and various classical methods, including GSA \cite{aiazzi2007improving}, CNMF \cite{yokoya2011coupled}, GLP \cite{aiazzi2006mtf}, MAPSMM \cite{eismann2004resolution}, and HySure \cite{simoes2014convex} is provided in Figure~\ref{fig:fused1}. In this figure, a small region of the original image is marked with a red box, and a zoomed-in view is shown in the lower-right corner. This close-up comparison clearly shows that HyDeFuse performs better than the other methods. The corresponding metrics are reported in Table~\ref{tab:fused_old}. 

\begin{figure}[t]
  \centering
  \begin{minipage}{0.24\textwidth}
      \centering
      \includegraphics[width=\linewidth]{output_Ground_truth_pavia_new_Setup.png} 
      \caption*{{(a) ground truth}} 
     
  \end{minipage} \hfill
  \begin{minipage}{0.24\textwidth}
      \centering
      \includegraphics[width=\linewidth]{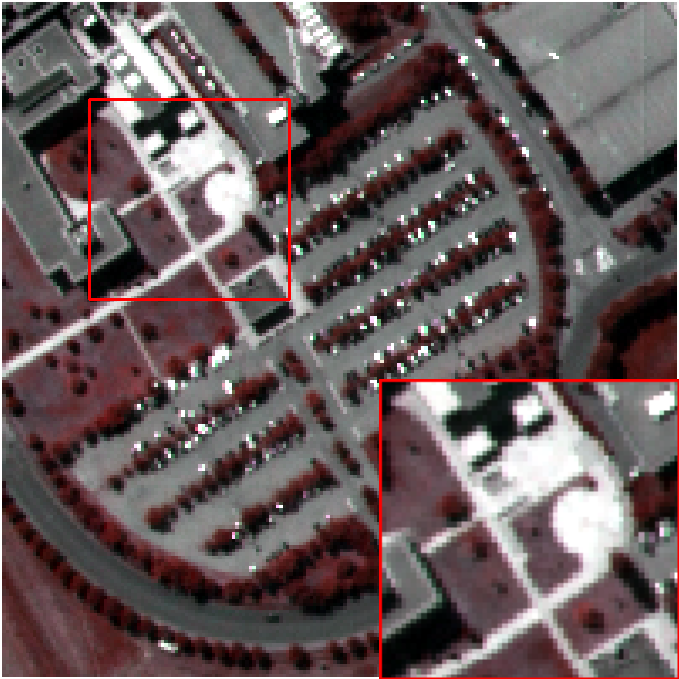} 
      \caption*{{(b) GTNN}} 
      
  \end{minipage} \hfill
  \begin{minipage}{0.24\textwidth}
      \centering
      \includegraphics[width=\linewidth]{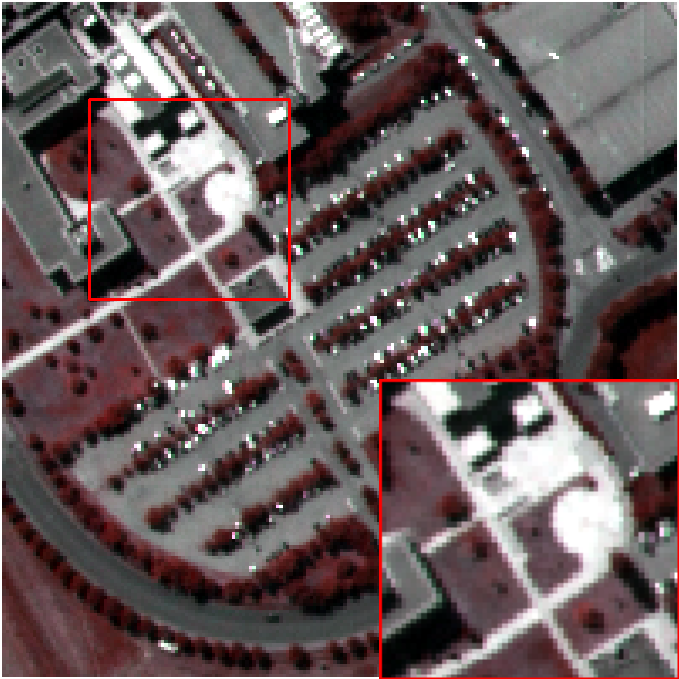} 
      \caption*{{(c) CTDF }} 
      
  \end{minipage} \hfill
  \begin{minipage}{0.24\textwidth}
      \centering
      \includegraphics[width=\linewidth]{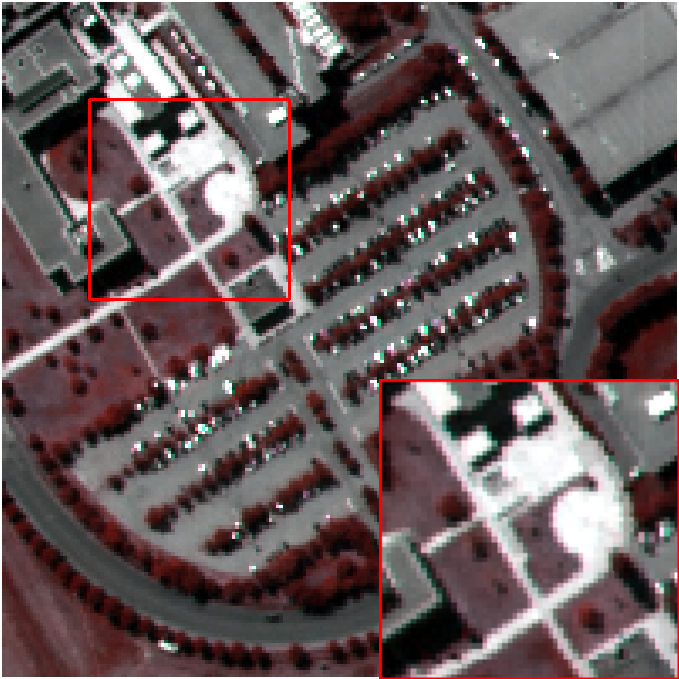} 
      \caption*{{(d) HyDeFuse}} 
      
  \end{minipage}
  \caption{Comparison of recent tensor-based methods on the Pavia dataset. The blur operator $\B$ is modeled using a Gaussian point spread function (PSF) with a radius of 7 and a standard deviation of 2. The corresponding performance metrics are listed in Table~\ref{tab:fused1}.}
  \label{fig:fused2}
\end{figure}

We next compare HyDeFuse with two recent methods: GTTN~\cite{dian2024hyperspectral} and CTDF~\cite{xu2024coupled}. The experiments are conducted on the Pavia dataset. A visual comparison is provided in Figure~\ref{fig:fused2}, while various performance metrics are presented in Tables~\ref{tab:fused1} and~\ref{tab:fused2}. We observe that our method, HyDeFuse, performs on par with these approaches. 

To further test its effectiveness and generality, we evaluate HyDeFuse under two different SNR settings—one used in the original GTTN and CTDF papers, and another used consistently in our comparisons.
Under high noise in both the HS and MS images, HyDeFuse clearly outperforms these recent methods, as shown in Table~\ref{tab:fused2}. This improvement is mainly due to the robustness of the CasKD denoiser, which adapts well to different noise levels. Some differences in performance may also be due to hyperparameter tuning, which could potentially improve the performance of competing methods.

In summary, HyDeFuse consistently outperforms classical fusion methods. Compared to recent methods such as GTTN and CTDF, it remains competitive under standard conditions and achieves significant improvements at higher noise levels, due to the adaptive capabilities of the CasKD denoiser.

\begin{table}[h]
\centering
\normalsize 
\begin{tabular}{|c|c|c c c c c|}
\hline

{Dataset} & {Methods} & {PSNR} & {RMSE} & {ERGAS} & {SAM} & {UIQI} \\
\hline
 & Bicubic & 25.05 & 0.0559 & 8.44 & 8.40 & 0.746 \\
 & MASMM & 28.63 & 0.0384 & 5.85 & 9.61 & 0.910 \\
 & GLP & 28.39 & 0.037 & 5.95 & 9.83 & 0.901 \\
Pavia & CNMF & 29.46 & 0.033 & 5.28 & 8.09 & 0.924 \\
 & GSA & 29.95 & 0.0318 & 5.03 & 9.98 & 0.918 \\
 & HySure & $\mathbf{34.12}$ & $\mathbf{0.020}$ & $\mathbf{3.02}$ & $\mathbf{5.39}$ & $\mathbf{0.965}$ \\
 & HyDeFuse & \underline{$\mathbf{35.89}$} & \underline{$\mathbf{0.016}$} & \underline{$\mathbf{2.39}$} & \underline{$\mathbf{3.61}$} & \underline{$\mathbf{0.980}$} \\
\hline
 & Bicubic & 23.05 & 0.0702 & 6.24 & 6.21 & 0.57 \\
 & MASMM & 25.01 & 0.0556 & 5.06 & 5.59 & 0.7761 \\
 & GLP & 24.36 & 0.059 & 5.43 & 6.35 & 0.751 \\
Paris & CNMF & 25.39 & 0.05 & 4.85 & 4.55 & 0.80 \\
 & GSA & 24.48 & 0.06 & 5.36 & 5.78 & 0.78 \\
 & HySure & $\mathbf{26.82}$ & $\underline{\mathbf{0.04}}$ & $\mathbf{4.13}$ & $\mathbf{3.17}$ & $\mathbf{0.839}$ \\
 & HyDeFuse & \underline{$\mathbf{27.24}$} & \underline{$\mathbf{0.04}$} & \underline{$\mathbf{3.93}$} & \underline{$\mathbf{2.81}$} & \underline{$\mathbf{0.846}$} \\
\hline
 & Bicubic & 29.02 & 0.035 & 7.49 & 7.30 & 0.54 \\
 & MASMM & 30.84 & 0.028 & 6.51 & 8.74 & 0.717\\
 & GLP & 30.20 & 0.030 & 6.80 & 8.31 & 0.698 \\
Chikusei & CNMF & 34.14 & 0.019 & 3.97 & 5.23 & 0.847 \\
 & GSA & 32.29 & 0.020 & 6.89 & 8.84 & 0.761 \\
 & HySure & $\mathbf{40.56}$ & $\mathbf{0.009}$ & $\mathbf{1.99}$ & $\mathbf{2.58}$ & $\mathbf{0.937}$ \\
 & HyDeFuse & \underline{$\mathbf{42.27}$} & \underline{$\mathbf{0.007}$} & \underline{$\mathbf{1.39}$} & \underline{$\mathbf{1.79}$} & \underline{$\mathbf{0.967}$} \\
\hline
\end{tabular}
\caption{Comparison of HyDeFuse with classical fusion methods across three datasets.}
\label{tab:fused_old}
\end{table}

\begin{table}[h]
\centering
\normalsize
\begin{tabular}{|c|c c c c c|}
\hline

{Method} & {PSNR} & {RMSE} & {ERGAS} & {SAM} & {UIQI} \\
\hline
CTDF & 38.49 & 0.013 & 1.94 & 3.67 & $\mathbf{0.9869}$ \\
GTNN & \underline{$\mathbf{41.58}$} & \underline{$\mathbf{0.008}$} & $\mathbf{1.37}$ & $\mathbf{2.39}$ & $0.9859$ \\
HyDeFuse & $\mathbf{40.79}$ & $\mathbf{0.009}$ & \underline{$\mathbf{1.33}$ }& \underline{$\mathbf{2.03}$ }& \underline{$\mathbf{0.994}$} \\
\hline
\end{tabular}
\caption{Quality metrics for Figure~\ref{fig:fused2}. In this experiment, \( \text{SNR}_h = 30 \, \text{dB} \) and \( \text{SNR}_m = 35 \, \text{dB} \).}
\label{tab:fused1}
\end{table}

\begin{table}[h]
\centering
\normalsize
\begin{tabular}{|c c c c c c|}
\hline

{Method} & {PSNR} & {RMSE} & {ERGAS} & {SAM} & {UIQI} \\
\hline
CTDF & 28.37 & 0.038 & 6.28 & 11.82 & 0.88 \\
GTNN & $\mathbf{31.53}$ & $\mathbf{0.027}$ & $\mathbf{4.60}$ & $\mathbf{8.15}$ & $\mathbf{0.9325}$ \\
HyDeFuse & \underline{$\mathbf{34.48}$} & \underline{$\mathbf{0.018}$} & \underline{$\mathbf{2.92}$} & \underline{$\mathbf{4.17}$} & \underline{$\mathbf{0.972}$} \\
\hline
\end{tabular}
\caption{Quality metrics evaluated on the Pavia dataset. The blurring operator $\B$ is modeled using a Gaussian point spread function (PSF) with a radius of 7 and a standard deviation of 2. In this experiment, $\text{SNR}_h = 20\, \text{dB}$ and $\text{SNR}_m = 20\, \text{dB}$.}
\label{tab:fused2}
\end{table}

 \section{Conclusion}  
 
We developed an iterative, denoising-based algorithm for hyperspectral fusion within the plug-and-play (PnP) framework. The proposed method combines strong empirical performance with provable convergence guarantees. This is achieved through the use of kernel denoisers, which provide effective noise suppression while maintaining mathematically tractable properties. In particular, we introduced CasKD, a cascaded kernel denoiser that captures both inter-band and intra-band correlations in hyperspectral data. CasKD outperforms previously reported kernel denoisers and performs on par with leading classical (non-trained) methods. We did not consider learned deep denoisers, as ensuring predictable mathematical behavior and convergence guarantees for such models remains fundamentally challenging.

To integrate the implicit CasKD regularizer with a model-based loss for hyperspectral fusion, we employed the classical proximal gradient descent (PGD) framework. We then established a rigorous operator-theoretic formulation of CasKD and the resulting fusion algorithm, HyDeFuse, providing a clear connection between the denoiser properties and the convergence behavior of the iterative scheme. Based on this formulation, we derived an unconditional guarantee of global linear convergence for HyDeFuse.

While this result ensures algorithmic stability and reliability, it does not by itself guarantee high-quality reconstructions. To validate the practical performance, we conducted extensive experiments on multiple hyperspectral datasets, showing that HyDeFuse consistently outperforms classical approaches and remains competitive with recent state-of-the-art methods. With suitable parameter settings, the PSNR stabilizes within 10–15 iterations. The underlying principles and convergence analysis can naturally extend to other denoising-based regularization models.

\section*{Data Availability Statement}
All data that support the findings of this study are included within the article. The code used
in this article will be made publicly available at 
\href{https://github.com/sagarkr1729/Contractive-Hyperspectral-Fusion}{https://github.com/sagarkr1729/Contractive-Hyperspectral-Fusion} 
upon acceptance of this manuscript.

\section{Appendix}
\label{appendix}

In this section, we prove~\Cref{prop:betasmooth}, \Cref{prop:G}, \Cref{prop:VandW}, \Cref{propfix}, and \Cref{lemmaT}.

\subsection{Proof of~\texorpdfstring{\Cref{prop:betasmooth}}{Cref(prop:betasmooth)}}
\label{betasmooth}

We first show that $\cK$ is self-adjoint on $\bbH$. Indeed, for any $\X_1, \X_2 \in \bbH$, we have
\begin{eqnarray*}
\big \langle \cK(\X_1), \X_2 \big \rangle_{\bbH} 
&=& \tr \big( (\mathbf{P}_1 \X_1 + \X_1 \mathbf{P}_2)^\top \X_2 \big) \nonumber \\
&=& \tr (\X_1^\top \mathbf{P}_1 \X_2) + \tr (\mathbf{P}_2 \X_1^\top \X_2) \nonumber \\
&=& \tr (\X_1^\top \mathbf{P}_1 \X_2) + \tr (\X_1^\top \X_2 \mathbf{P}_2) \nonumber \\
&=& \big \langle \X_1, \cK(\X_2) \big \rangle_{\bbH},
\label{eq:selfadjoint_eqnarray}
\end{eqnarray*}
where the cyclic property of trace and the symmetry of $\mathbf{P}_1$ and $\mathbf{P}_2$ are used. Hence, $\cK$ is self-adjoint.

We next show that $\cK$ is $\beta$-Lipschitz, where 
\begin{equation}
\label{eqL:formula-beta}
\beta = \sigma_{\max}(\A)^2 + \lambda \, \sigma_{\max}(\E \R)^2.
\end{equation}
From~\eqref{eq:K} and the triangle inequality, 
\begin{equation*}
\label{eq:Kbound_eqnarray}
\|\cK(\X)\|_{\bbH}  \leqslant \|\mathbf{P}_1 \X\|_{\bbH} + \|\X \mathbf{P}_2\|_{\bbH}.
\end{equation*}
Let $\mathbf{P}_1 = \mathbf{U} \mathbf{\Sigma} \mathbf{V}^\top$ be the SVD of $\mathbf{P}_1$, where $\mathbf{U}$ and $\mathbf{V}$ are orthogonal, and $\mathbf{\Sigma} = \mathrm{diag}(\sigma_1, \dots, \sigma_r)$ are the singular values. Since orthogonal matrices preserve the Frobenius norm, 
\begin{equation*}
\|\mathbf{P}_1 \X\|_{\bbH} = \|\mathbf{U} \mathbf{\Sigma} \mathbf{V}^\top \X\|_{\bbH} = \|\mathbf{\Sigma} \mathbf{Y}\|_{\bbH}, \qquad \mathbf{Y} := \mathbf{V}^\top \X.
\end{equation*}
Because $\mathbf{\Sigma}$ is diagonal,
\begin{equation*}
\|\mathbf{\Sigma} \mathbf{Y}\|_{\bbH}^2 = \sum_{i,j} \sigma_i^2 |\Y_{ij}|^2 \leqslant \sigma_{\max}(\mathbf{P}_1)^2 \sum_{i,j} |\Y_{ij}|^2 = \sigma_{\max}(\mathbf{P}_1)^2 \|\mathbf{Y}\|_{\bbH}^2.
\end{equation*}
Since $\mathbf{V}$ is orthogonal, $\|\mathbf{Y}\|_{\bbH} = \|\X\|_{\bbH}$, giving
\begin{equation*}
\|\mathbf{P}_1 \X\|_{\bbH}  \leqslant \sigma_{\max}(\mathbf{P}_1) \|\X\|_{\bbH}.
\end{equation*}
Similarly, taking transposes,
\begin{equation*}
\|\X \mathbf{P}_2\|_{\bbH} = \|\mathbf{P}_2^\top \X^\top\|_{\bbH}  \leqslant \sigma_{\max}(\mathbf{P}_2) \|\X^\top\|_{\bbH} = \sigma_{\max}(\mathbf{P}_2) \|\X\|_{\bbH}.
\end{equation*}
Combining these bounds, we obtain
\begin{equation}
\label{eq:beta-lip}
\|\cK(\X)\|_{\bbH}  \leqslant \beta \|\X\|_{\bbH}, \quad \beta:=\sigma_{\max}(\mathbf{P}_1) + \sigma_{\max}(\mathbf{P}_2).
\end{equation}
Hence, $\cK$ is $\beta$-Lipschitz. Substituting $\mathbf{P}_1$ and $\mathbf{P}_2$ from~\eqref{eq:K} yields the expression for $\beta$ in~\eqref{eqL:formula-beta}.

\subsection{Proof of Proposition~\ref{prop:G}}
\label{pf:G}

From~\eqref{eq:gradK}, we have
\begin{equation}
\label{eq:grad-K}
\nabla \! \ell(\X_1) - \nabla \!\ell(\X_2)  =  \cK(\X_1) - \cK(\X_2) = \cK(\X_1-\X_2).
\end{equation}
Hence, by~\eqref{eq:beta-lip}, $\nabla \! \ell$ is $\beta$-smooth. Moreover, since $\ell$ is convex, the Baillon–Haddad theorem~\cite{bauschke2011convex} gives
\begin{equation}
\label{eq:baillon}
 \langle  \nabla \! \ell(\X_1) - \nabla \!\ell(\X_2) , \X_1 - \X_2\ \rangle_{\bbH}  \geqslant  \frac{1}{\beta}\| \nabla \! \ell(\X_1) - \nabla \!\ell(\X_2)\|_{\bbH}^2  \qquad (\X_1,\X_2 \in \bbH) . 
\end{equation}
Substituting \eqref{eq:grad-K}, we obtain
\begin{equation}
\label{eq:BH1}
 \langle  \cK(\X) , \X \ \rangle_{\bbH}  \geqslant  \frac{1}{\beta}\| \cK(\X) \|_{\bbH}^2  \qquad (\X \in \bbH) . 
\end{equation}
From \eqref{eq:form}, we can write
\begin{equation*}
\| \cG(\X)\|_{\bbH}^2 =  \| \X \|_{\bbH}^2 + \gamma^2 \| \cK(\X)\|_{\bbH}^2 - 2 \gamma \langle  \X, \cK(\X)  \ \rangle_{\bbH}.
\end{equation*}
Substituting~\eqref{eq:BH1} gives
\begin{equation}
\label{eq:balancegamma}
\| \cG(\X)\|_{\bbH}^2 \leqslant  \| \X \|_{\bbH}^2 - \gamma \left(\frac{2}{\beta}- \gamma \right) \| \cK(\X)\|_{\bbH}^2 \leqslant  \| \X \|_{\bbH}^2.
\end{equation}
Hence, if $0 < \gamma \leqslant 2/\beta$, we have $\| \cG(\X)\|_{\bbH} \leqslant \|\X\|_{\bbH}$. Since $\cG \in \cL(\bbH)$, it follows that $\cG$ is nonexpansive.

To complete the proof of Proposition~\eqref{prop:G}, we must show that $-1$ is not an eigenvalue of $\cG$. Indeed, if $-1 \in \sigma(\cG)$, then there exists a nonzero $\overline{\X} \in \bbH$ such that
\begin{equation*}
\cG(\overline{\X})= \overline{\X} -  \gamma \cK(\overline{\X}) = -\overline{\X},
\end{equation*}
However, this implies  $2\overline{\X}=\gamma \cK(\overline{\X})$, so that if $0 < \gamma < 2/\beta$, we have from \eqref{eq:beta-lip} that
\begin{equation*}
2 \|\overline{\X}\|_{\bbH} = \gamma \|\cK(\overline{\X})\|_{\bbH}< \frac{2}{\beta}\|\cK(\overline{\X})\|_{\bbH} \leqslant 2 \|\overline{\X}\|_{\bbH},
\end{equation*}
which is a contradiction. Therefore, we must have \( -1 \notin \sigma(\cG) \).

\subsection{Proof of Proposition~\ref{prop:VandW}}
\label{pf:VandW}

Recall that $\cV$ is the bandwise denoiser defined in~\eqref{eq:defV}, while $\cW$ is the high-dimensional denoiser defined in~\eqref{eq:defW}. The key distinction is that $\cV$ applies a separate denoiser $\W^{(b)}$ to each spectral band (see~\eqref{eq:defWb}), whereas $\cW$ uses a single shared denoiser $\W$ across all bands (see~\eqref{eq:defbW}).

To show that $\cV$ is self-adjoint, we need to verify that
\begin{equation*}
\big \langle \cV(\X), \Y \big \rangle_{\bbH}= \big \langle \X, \cV(\Y) \big \rangle_{\bbH} \qquad (\X, \Y \in \bbH).
\end{equation*}
Recall that $\bbH$ is the space of matrices of size $N_m \times L_s$. It will be convenient to denote by $\x_b$ and $\y_b$ the $b$-th columns (vectors in $\Re^{N_m}$) of $\X$ and $\Y$, respectively. With this notation, we can rewrite~\eqref{eq:innerprod} as
\begin{equation*}
 \langle \X, \Y \rangle_{\bbH} = \sum_{b=1}^{L_s} \x_b^\top \y_b.
\end{equation*}
As the $b$-th band of $\cV(\X)$ is $\W^{(b)} \x_b$ and $\W^{(b)}$ is symmetric~(Proposition~\ref{prop:WV}), we have
\begin{equation*}
\big \langle \cV(\X), \Y \big \rangle_{\bbH}=  \sum_{b=1}^{L_s} (\W^{(b)} \x_b)^\top \y_b = \sum_{b=1}^{L_s} \x_b^\top (\W^{(b)} \y_b) = \big \langle  \X, \cV(\Y) \big \rangle_{\bbH}.
\end{equation*}
The verification for $\cW$ is identical. 

Since $\W^{(b)}$ and $\W$ are stochastic and symmetric, it is straightforward to verify that they are nonexpansive. Consequently, the associated operators $\cV$ and $\cW$ are also nonexpansive. Indeed, for any $\X \in \bbH$, we have 
\begin{equation*}
\|\cV(\X)\|^2_{\bbH} =  \sum_{b=1}^{L_s} \| \W^{(b)} \x_b\|^2_2 \leqslant \sum_{b=1}^{L_s} \|\x_b\|^2_2 = \|\X\|^2_{\bbH}.
\end{equation*}
where $\|\cdot \|_2$ is the $\ell_2$ norm on $\Re^{N_m}$. The argument for $\cW$ is identical.

We know that a self-adjoint operator has real eigenvalues~\cite{meyer2023matrix}. Moreover, since $\cW$ has been shown to be nonexpansive, it follows that if $\lambda \in \sigma(\cW)$, then $|\lambda| \leqslant 1$, i.e., $\sigma(\cW) \subset [-1,1]$. Further, from Proposition~\ref{prop:WV}, $\W$ is symmetric and positive semidefinite. Hence, for any $\X \in \bbH$,
\begin{equation*}
\langle \cW(\X), \X \rangle_{\bbH} = \sum_{b=1}^{L_s} (\W \x_b)^\top \x_b = \sum_{b=1}^{L_s} \x_b^\top \W \x_b \geqslant 0.
\end{equation*}
This shows that all eigenvalues of $\cW$ are nonnegative, and therefore $\sigma(\cW) \subset [0,1]$.

Finally, we show that if $\cW(\X)=\X$, then each band $\x_b$ must be a multiple of $\e$, the all-ones vector in $ \Re^{N_m}$. This follows directly from the fact that $1$ is a simple eigenvalue of $\W$. 
Indeed, the equation $\cW(\X)=\X$ translates to $\W\x_b = \x_b$ for all $b$, i.e., $\x_b$ is an eigenvector with eigenvalue $1$. Since $\W$ is stochastic and irreducible~(Proposition~\ref{prop:WV}), the Perron-Frobenius theorem tells us that $1$ must be a simple eigenvalue of $\W$~\cite{meyer2023matrix}. 
Consequently, we must have $\x_b = c_b \e$ for some $c_b \in \Re$. This establishes \eqref{eq:F} and completes the proof of Proposition~\ref{prop:VandW}.

\subsection{Proof of Proposition~\ref{propfix}}
\label{pf:propfix}

From Proposition~\ref{prop:VandW}, we know that $\fix(\cW)$ consists of matrices $\F$ of the form~\eqref{eq:F}. Therefore, to establish the stated result, it suffices to show that if $\F$ is of the form~\eqref{eq:F} and $\F \neq \0$, then $\F \notin \fix(\cG)$. 

We will show, by contradiction, that under this condition $\F \notin \fix(\cG)$. Suppose that $\cG(\F)=\F$. Then, from \eqref{eq:form}, we have $\cK(\F)=\0$, i.e., 
\begin{equation*}
 \mathbf{P}_1 \F + \F \mathbf{P}_2 = \0,
\end{equation*}
where $ \mathbf{P}_1$ and $ \mathbf{P}_2$ are given by~\eqref{eq:K}. In particular, this means
\begin{equation*}
0=\langle \F,  \mathbf{P}_1 \F + \F \mathbf{P}_2 \rangle_{\bbH}= \langle \F,  \mathbf{P}_1 \F\rangle_{\bbH} + \langle \F^\top\!,  \mathbf{P}_2 \F^\top \rangle_{\bbH}.
\end{equation*}
Since $\mathbf{P}_1 $ and $\mathbf{P}_2$ are positive semidefinite, we know that the two terms on the right 
are nonnegative. Therefore, we must have
\begin{equation*}
\langle \F,  \mathbf{P}_1 \F\rangle_{\bbH} =0 \quad \mbox{and} \quad \langle \F^\top\!,  \mathbf{P}_2 \F^\top \rangle_{\bbH}=0.
\end{equation*}
We just need the first relation. Substituting~\eqref{eq:form} and noting that $\mathbf{P}_1=\A^\top\!\A$, we get
\begin{equation*}
0= \langle \F,  \mathbf{P}_1 \F\rangle_{\bbH}= \tr(\F^\top\!  \mathbf{P}_1 \F)= \big(c_1^2+\cdots+c_{L_s}^2 \big)  \| \A\e\|_2^2,
\end{equation*}
where $\|\cdot \|_2$ is the $\ell_2$ norm on $\Re^{N_m}$. Now, since $\F \neq \0$, we have  $c_j \neq 0$ for some $1 \leqslant j \leqslant L_s$. 
Hence, it follows that $\A\e = \0$. 

Recall that $\A=\S\B$, where $\S$ and $\B$ are sampling and blur operators. Since $\B\e=\e$, i.e., blurring a constant-intensity image produces the same image, we obtain  $\0=\A\e=\S(\B\e)=\S\e$. 
This leads to a contradiction, as sampling $\e$ cannot yield the zero vector.

\subsection{Proof of Lemma~\ref{lemmaT}}
\label{pf:lemmaT}

Since $\cT$ is a self-adjoint operator on the inner product space $\bbH$, it can be diagonalized with respect to an orthonormal basis~\cite{meyer2023matrix}. Specifically, if $d$ is the dimension of $\bbH$, then we can find $\lambda_1,\ldots, \lambda_d \in \Re$ and an orthonormal basis $\H_1,\ldots,\H_d$ of $\bbH$ such that, for any $\X \in \bbH$,
\begin{equation}
\label{eq:XTX}
\X = \sum_{j=1}^d \langle \X, \H_j \rangle_{\bbH} \ \H_j \quad \mbox{and}  \quad \cT(\X) = \sum_{j=1}^d \lambda_j \langle \X, \H_j \rangle_{\bbH}  \ \H_j.
\end{equation}
In particular,
\begin{equation}
\label{eq:parseval}
\| \X  \|^2_{\bbH}= \sum_{j=1}^d \langle \X, \H_j \rangle_{\bbH}^2 \quad \mbox{and}  \quad \| \cT(\X) \|^2_{\bbH}= \sum_{j=1}^d \lambda^2_j \langle \X, \H_j \rangle_{\bbH}^2.
\end{equation}
Now, from~\eqref{eq:parseval}, we have that $\| \cT(\X) \|_{\bbH}=\| \X \|_{\bbH}$ if and only if
\begin{equation*}
\sum_{j=1}^d (1- \lambda^2_j)\langle \X, \H_j \rangle_{\bbH}^2 =0.
\end{equation*}
Since $\lambda_j \in (-1,1]$ by assumption, it follows that either  $\lambda_j = 1$ or $\langle \X, \H_j \rangle_{\bbH} = 0$. In both cases, it follows from~\eqref{eq:XTX} that $\cT(\X) = \X$. This completes the proof of Lemma~\ref{lemmaT}.

\bibliographystyle{ieeetr}
\bibliography{refs}

\end{document}